\pdfoutput=1

\documentclass[11pt,twoside,a4paper,cmspaper,final,collab]{cms-tdr}
\def\svnVersion{5b40570-D}\def\svnDate{2020/03/31}\def\cmsCernNoTag{CERN-EP-2019-280}\def\cmsCernDate{\today}\def\cmsMessage{"Published in the Journal of High Energy Physics as \href{http://dx.doi.org/10.1007/JHEP05(2020)052}{\doi{10.1007/JHEP05(2020)052}.}"}

\begin{document}\cmsNoteHeader{EXO-18-013}

\hyphenation{had-ron-i-za-tion}
\hyphenation{cal-or-i-me-ter}
\hyphenation{de-vices}
\RCS$HeadURL: svn+ssh://svn.cern.ch/reps/tdr2/papers/EXO-18-013/trunk/EXO-18-013.tex $
\RCS$Id: EXO-18-013.tex 491846 2019-03-14 21:27:09Z hoepfner $

\providecommand{\cmsTable}[1]{\resizebox{\textwidth}{!}{#1}}

\newcommand{\mlstar}{\ensuremath{M_{\ell^*}}\xspace }
\newcommand{\lstar}{\ensuremath{\ell^*}\xspace }
\newcommand{\estar}{\ensuremath{\Pe^*}\xspace }
\newcommand{\mustar}{\ensuremath{\Pgm^*}\xspace }
\newcommand{\llZ}{\ensuremath{\ell \ell \PZ}\xspace}
\newcommand{\Pj}{\HepParticle{j}{}{}}
\newcommand{\llg}{\ensuremath{\ell\ell\gamma}\xspace}
\newcommand{\eeee}{\ensuremath{\Pe\Pe^* \to  \Pe\Pe\PZ  \to 4\Pe}\xspace}
\newcommand{\eemm}{\ensuremath{\Pe\Pe^* \to  \Pe\Pe\PZ \to  2\Pe 2\Pgm}\xspace}
\newcommand{\eejj}{\ensuremath{\Pe\Pe^* \to \Pe\Pe\PZ  \to  2\Pe 2\Pj }\xspace}
\newcommand{\mmmm}{\ensuremath{\Pgm\Pgm^* \to \Pgm\Pgm\PZ  \to 4\Pgm}\xspace}
\newcommand{\mmee}{\ensuremath{\Pgm\Pgm^* \to \Pgm\Pgm\PZ  \to 2\Pgm 2\Pe}\xspace}
\newcommand{\mmjj}{\ensuremath{\Pgm\Pgm^* \to \Pgm\Pgm\PZ  \to 2\Pgm 2\Pj}\xspace}

\newcommand{\lljj}{\ensuremath{\ell\ell jj}\xspace}
\newcommand{\llll}{\ensuremath{\ell\ell^* \to \ell\ell\PZ  \to 4\ell }\xspace}
\newcommand{\Mmin}{\ensuremath{M_{\text{min}}}\,  }
\newcommand{\Mmax}{\ensuremath{M_{\text{max}}}\,  }
\newcommand{\sqrts}{\ensuremath{\sqrt{s}}}

\newlength\cmsTabSkip\setlength{\cmsTabSkip}{1ex}

\cmsNoteHeader{EXO-18-013}
\title{Search for an excited lepton that decays via a contact interaction
to a
lepton and two jets in proton-proton collisions at $\sqrts = 13\TeV$}

\date{\today}

\abstract{
Results are presented from a search for events containing an excited lepton (electron or muon)
produced in association with an ordinary lepton of the same flavor
and decaying to
a lepton and two hadronic jets.
Both the production and the decay of
the excited leptons are assumed to occur via a contact interaction with
a characteristic energy scale $\Lambda$.
The branching fraction for the decay mode under study
increases
with the mass of the excited lepton and is the most sensitive channel for very heavy excited
leptons.
The analysis uses a sample of proton-proton collisions collected by the CMS experiment at the LHC
at $\sqrts = 13\TeV$, corresponding to an integrated luminosity of 77.4\fbinv.
The four-body invariant mass of the two lepton plus two jet
system is used as the primary discriminating variable. No significant excess of events beyond the
expectation for standard
model processes is observed. Assuming that $\Lambda$ is equal to the mass of the excited leptons, excited electrons and
muons
with masses below 5.6 and 5.7\TeV, respectively, are excluded at 95\%
confidence level. These are the best limits to date.
}

\hypersetup{%
pdfauthor={CMS Collaboration},%
pdftitle={Search for an excited lepton that decays via a contact interaction to a lepton and two jets in pp collisions at sqrt(s) = 13 TeV},%
pdfsubject={CMS},%
pdfkeywords={CMS, physics, BSM, excited leptons, compositeness}}

\maketitle

\section{Introduction}
\label{sec:intro}

The standard model (SM) of particle physics accurately describes a broad range of observations, but it does not provide
an
explanation for many of its own features. Among the most prominent of these features is the existence of
three fermion generations,
encompassing both leptons and quarks.
Attempts to explain the observed generation structure have led to a class of models postulating that quarks and
leptons are composite objects that consist of more fundamental
constituents~\cite{Pati:1975md,compositeness1, Eichten:1983hw, harari,
ssc-physics, Baur90,   Greenberg:1974qb, Greenberg:1980ri, Biondini:2019tcc}.
In these models, the fundamental constituents are bound by an asymptotically free gauge interaction that
becomes strong below a
characteristic scale $\Lambda$.
Such compositeness models predict the existence of
excited states of quarks ($\PQq^{*}$) and leptons (\lstar) at the characteristic scale of the new binding interaction.
Since these excited fermions couple to ordinary SM fermions, they could be produced via contact interactions (CI) in
collider experiments,
with subsequent decays to SM fermions through the gauge interactions (GIs),
via the emission of a \PW or \PZ boson or a photon ($\gamma$), or via a CI to other fermions.

\begin{figure}[bh]
\centering
\includegraphics[width=0.5\textwidth]{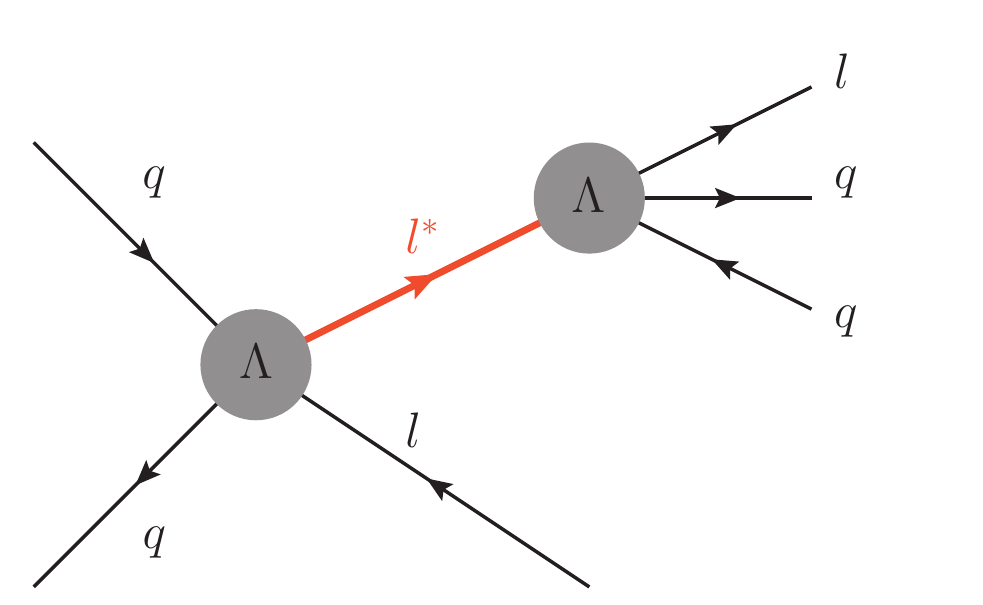}
\caption{Feynman diagram for the production of an excited lepton in association with an SM lepton in a hadron
collider. The excited lepton decays via a contact interaction to one SM lepton and two resolved jets, which result
from the
hadronization of the quarks.
}
\label{fig:introduction:feynman}
\end{figure}

Searches at the CERN LEP~\cite{Buskulic:1996tw, Abreu:1998jw, Abbiendi:1999sa, Achard:2003hd},
DESY HERA~\cite{H1estar}, and the Fermilab
Tevatron~\cite{CDFestar, cdfmu, d0, D0estar} have found no evidence for excited
leptons. At  the CERN Large  Hadron Collider (LHC), previous searches  performed  by the
ATLAS~\cite{atlas-limit_new, Aad:2016giq, Aaboud:2019zpc}
and CMS collaborations~\cite{cms-limit_new, EXO-14-015}
have also found no evidence of excited  leptons.
The maximum
$\Lambda$ value excluded experimentally corresponds to 25\TeV
(for $\mlstar\approx1\TeV$)
from a recent CMS \llg analysis
using proton-proton (pp) collision data at $\sqrts = 13\TeV$,
corresponding to an integrated luminosity of 35.9\fbinv~\cite{EXO-18-004}.
The same analysis excluded excited electrons and muons with masses below 3.7 and 3.8\TeV,
respectively,
for $\mlstar = \Lambda$.

In this paper, a search for excited leptons (\estar and \mustar) is presented.
This analysis focuses on the production of two same-flavor leptons of which one is excited and the other
one corresponds to an SM lepton. The excited lepton
is assumed to decay subsequently via a CI to an SM lepton of the same flavor
and a quark
pair, yielding two jets.
The process is illustrated in Fig.~\ref{fig:introduction:feynman}.

The data used for this analysis
were recorded with the CMS detector
in pp collisions at $\sqrts=13\TeV$ during 2016 and 2017 and correspond to integrated luminosities of
35.9 and 41.5\fbinv, respectively, for a total of 77.4\fbinv.

\section{The CMS detector}
\label{sec:detector}

The central feature of the CMS apparatus is a superconducting solenoid of 6\unit{m} internal diameter, providing a
magnetic field of 3.8\unit{T}.
Within the solenoid volume are a silicon
pixel and strip tracker, a lead tungstate crystal electromagnetic calorimeter (ECAL), and a brass and scintillator hadron
calorimeter (HCAL),
each composed of a barrel and two endcap sections.
Extensive forward calorimetry complements the coverage provided by the barrel and endcap detectors.

The silicon tracker measures charged particles within the pseudorapidity range $\abs{\eta}<2.5$. It consists of silicon
pixel and silicon strip detector modules.
The electromagnetic calorimeter consists of 75\,848 lead tungstate crystals that provide coverage in pseudorapidity $\abs{\eta}<1.48$
in a barrel region and $1.48<\abs{ \eta }<3.00$ in two endcap regions.
The ECAL energy resolution for electrons with a transverse momentum $\pt\approx45\GeV$ from $\PZ \to \Pe \Pe$ decays is
better than 2\% in the central region of the ECAL barrel $(\abs{\eta}<0.8)$, and
is between 2 and 5\% elsewhere~\cite{EGM-14-001}.
For high energies, which are relevant for this analysis, the electron energy resolution slightly improves~\cite{EXO-12-061}.

Muons are
measured in gas-ionization detectors embedded in the steel flux-return yoke outside the solenoid,
in the pseudorapidity range $\abs{\eta}<2.4$. Detection is provided using three technologies: drift tubes (DTs),
cathode strip chambers (CSCs),
and resistive-plate chambers (RPCs).
While the barrel region of $\abs{\eta} \leq1.1$ is instrumented with DTs and RPCs, the endcaps
($1.1<\abs{\eta}<2.4$)
are equipped with CSCs and RPCs.
A muon from the interaction point will cross four layers of muon chambers, interleaved with steel forming the
return yoke of the magnetic field.
Every chamber provides reconstructed hits
on several detection planes, which are then combined into local track segments, forming the basis of muon
reconstruction inside the muon system.
Matching muons to tracks measured in the silicon tracker results in a relative transverse momentum resolution, for
muons with \pt up to 100\GeV,
of 1\% in the barrel
and 3\% in the endcaps. The \pt resolution in the barrel is better than 7\% for muons with \pt up to
1\TeV~\cite{MuonPerformance13TeV}.

A particle-flow algorithm~\cite{Sirunyan:2017ulk} aims to reconstruct and identify
each individual particle in an event, with an
optimized
combination of information from the various elements of the CMS detector. The energy of photons is obtained from the ECAL measurement. The
energy of electrons is determined from a combination of the electron momentum at the primary interaction vertex as determined by the
tracker, the energy of the corresponding ECAL cluster, and the energy sum of all bremsstrahlung photons spatially compatible with
originating from the electron track. The muon \pt is obtained from the curvature of the corresponding track. The energy of charged
hadrons is determined from a combination of their momentum measured in the tracker and the matching ECAL and HCAL energy deposits,
corrected for zero-suppression effects and for the response function of the calorimeters to hadronic showers. Finally, the energy of
neutral hadrons is obtained from the corresponding corrected ECAL and HCAL energies.

The CMS experiment has a two-level trigger
system~\cite{Khachatryan:2016bia}.
The level-1 trigger, composed of custom hardware processors, selects events of interest using information from the calorimeters
and muon detectors and reduces the readout rate from the 40\unit{MHz} bunch-crossing frequency
to a maximum of 100\unit{kHz}. The software based high-level trigger uses the full event
information, including that from the inner tracker, to reduce the event rate to the 1\unit{kHz} that is
recorded.

A more detailed description of the CMS detector can be found in Ref.~\cite{Chatrchyan:2008zzk}.

\section{Model and signal simulation}
\label{sec:model}

\begin{figure}[btp]
\centering
   \includegraphics[width=0.45\textwidth]{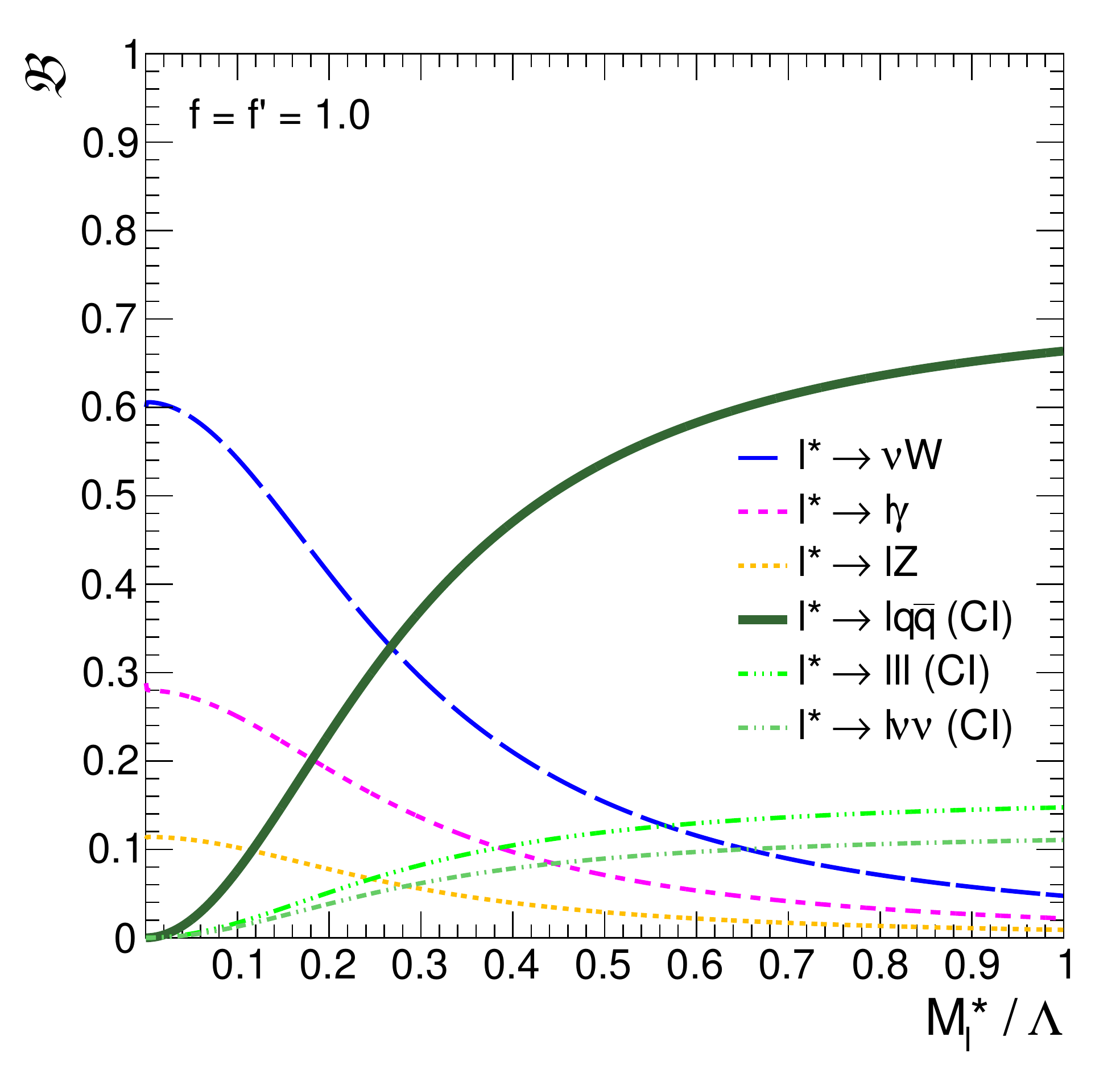}
   \includegraphics[width=0.45\textwidth]{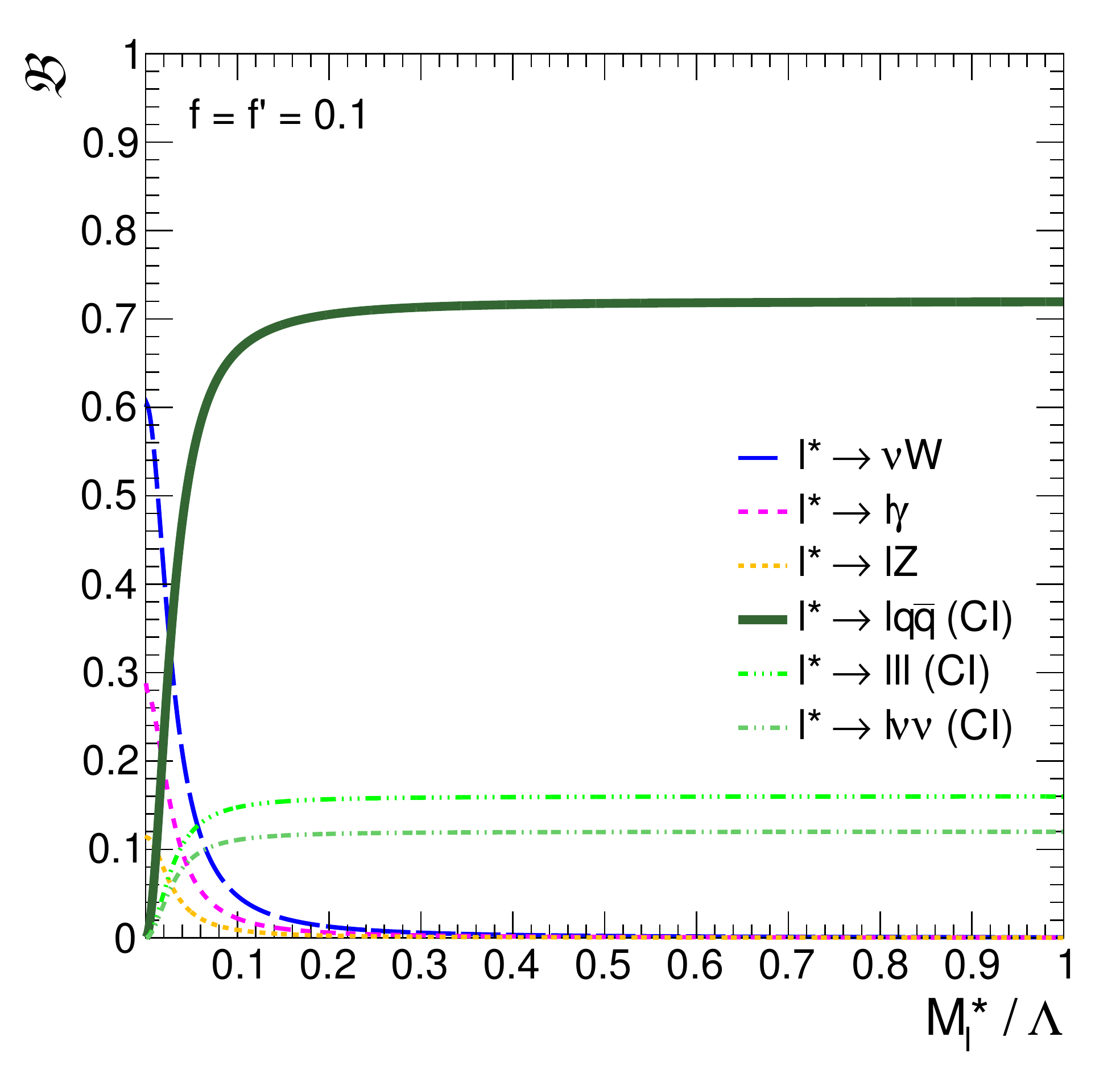}
   \caption{Branching fractions, $\mathcal{B}$,
of excited lepton decay channels as a function of the ratio of
the excited lepton mass (\mlstar) and
compositeness scale ($\Lambda$) for fixed values of the model parameters $f=f'$, which
represent the
couplings of excited leptons to SM particles.
The branching fraction
calculation is based on Ref.~\cite{Baur90}.
The contact interaction decay to one lepton and two jets, subject of this analysis, is dominating the
region of high \mlstar/$\Lambda$.
Couplings $f$ and $f'$ are assumed to be equal to 1 in the left graph, and 0.1 in the right graph.
}
\label{fig:br}
\end{figure}

Excited leptons can be produced at a pp collider by a CI that is described by an effective four-fermion Lagrangian
\begin{equation}
   \mathcal{L}_{\text{CI}} = \frac{g_{*}^2}{\Lambda^2} \frac{1}{2} j^{\mu} j_{\mu},
\end{equation}
with the coupling $g_{*}^2$ chosen to correspond to $4\pi$,
fermion current $j_{\mu} = \overline{\psi}\gamma_{\mu}\psi$,
and $\Lambda$ representing the compositeness scale~\cite{Baur90}.
The excited lepton, \lstar, can decay to an SM lepton via a CI $\lstar \to \ell \psi \overline{\psi}$,
where $\psi$ is a fermion, or through the mediation of a gauge boson.
This analysis complements other searches for excited leptons in the \llg or \llZ channels
and is most sensitive at large values of \mlstar and/or $\Lambda$.
Figure~\ref{fig:br}
illustrates
the rapid increase of the branching fraction of the CI decay
to a lepton plus two jets as a function of $\mlstar / \Lambda$.

The GI transition of excited leptons to SM leptons is described by
\begin{equation}
   \mathcal{L}_{\text{GI}} = \frac{1}{2\Lambda} \overline{\psi}_R^\star  \sigma^{\mu\nu} \left (
g f \frac{\boldsymbol{\tau}}{2}
\boldsymbol{W}_{\mu\nu} + g' f' \frac{Y}{2} B_{\mu\nu} \right ) \psi_L + \text{h.c.},
\end{equation}
where $g$ and $g^{\prime}$ denote the couplings to the hypercharge and
the weak isospin current and
$f$ and $f'$
effectively
scale the energy scale of the interaction $\Lambda$ with respect to the CI~\cite{Baur90}.
Other quantities are
$\boldsymbol{W}_{\mu\nu}$ and $B_{\mu\nu}$ denoting the field strength tensors of the
$\text{SU}(2)$ and $\text{U}(1)$ gauge
fields
with their generators $\boldsymbol{\tau}$ and $Y$.
The partial width of the CI decay is independent of $f$ and $f'$.
Interference between GI and CI transitions is not considered~\cite{Baur90}.
Since the exact values for $f$ and $f'$
are not known, searches for excited leptons are
typically
interpreted for two extreme values: $f = f' = 1$ and $f = -f' = 1$.
Gauge-interaction decays via photon emission are forbidden for $f = -f^\prime$.
Figure~\ref{fig:br} shows the variation of the branching fractions of the different 
decay channels as a function of \mlstar/$\Lambda$, illustrated in Fig.~\ref{fig:br} 
(left) for couplings of 1.0.  
Couplings below unity are also possible and can have a significant impact
on the branching fractions, as seen in Fig.~\ref{fig:br} (right)
for couplings of 0.1.
Weaker couplings reduce widths for the gauge-interaction decays while enhancing
the CI decays. A range of couplings is included in the interpretation of the
experimental results.

The production and decay of excited leptons are simulated at leading-order precision
with \PYTHIA 8.212~\cite{Sjostrand:2014zea}
using the NNPDF2.3~\cite{ball_parton_2013} (in 2016)
and
NNPDF3.1~\cite{Ball:2017nwa} (in 2017)
parton distribution functions (PDF) and tunes
CUETP8M1 (2016) and CP5 (2017)
for the underlying event, respectively~\cite{GEN-17-001}.
Simulated samples are generated for
\lstar masses of 200, 500, 750 and 1000\GeV and from 1 to 7\TeV in steps of
0.5\TeV.
In the simulation, all couplings are set to unity and $\Lambda$ to 10\TeV.
The specific choice of $\Lambda$ in the simulation scales the overall cross section
for the process, while leaving the kinematic variables unaffected.
Hence, the simulated events can be reweighted in order to represent different choices of the couplings
and  $\Lambda$. While the resonance width depends on $\Lambda$, it is smaller than the detector resolution for
all values used
in this analysis. For the example of a 5\TeV signal, the width is of order 15\% of the resolution.

At $\Lambda = 10\TeV$ and $f=1$
the product of the cross section and branching fraction for the channel under study ranges from
0.224\unit{fb} ($\mlstar = 200\GeV$), 1.115\unit{fb} ($\mlstar = 1\TeV$),
1.07 $\times 10^{-2}\unit{fb}$ ($\mlstar = 5\TeV$) to 6.91 $\times 10^{-4}\unit{fb}$ ($\mlstar = 7\TeV$).
For a high-mass  \lstar signal, each of the decay products will have a \pt of approximately
1\TeV, and will tend to be in the
central part of the detector.

\section{Event reconstruction}
\label{sec:objects}

The candidate vertex with the largest value of summed physics-object $\pt^2$ is taken to be the
primary $\Pp\Pp$ interaction vertex.
Here the physics objects are the jets, clustered using the
jet finding algorithm~\cite{Cacciari:2008gp,Cacciari:2011ma}
with the tracks assigned to the candidate vertices as inputs, and the associated
missing transverse momentum, taken as the negative vector \pt sum of these jets.
In order to suppress background from cosmic ray muons, an event must have
at least one primary vertex, with at least five associated well-reconstructed tracks
that are
within 24\cm in the longitudinal and 2\cm in the transverse direction from the nominal
interaction point.

Electrons are reconstructed as ECAL clusters that are matched to a central track and their
identification has been optimized for high-\pt values~\cite{ZprimeFull2016}.
In order to differentiate between electrons and photons, the properties of the track matched to the
calorimeter measurement must be consistent with those of an electron
originating from the primary vertex.
Specifically, there must be $\leq$1 hit missing in the innermost tracker layers, and the transverse distance
to the primary vertex must be less than 2\unit{mm} (barrel) or less than 5\unit{mm} (endcap).
Electron candidates are required to have an electron-like shower shape, and to be within the
acceptance region of the barrel ($\abs{\eta}<1.44$)
or the endcaps ($1.56<\abs{\eta}<2.50$), thus avoiding
the transition region between the barrel and the endcap parts of the ECAL.
The corresponding track matched to the ECAL cluster must be consistent with a particle
originating from the nominal interaction point.
The associated energy in the HCAL around the electron direction must be less than 5\%
of the reconstructed energy
of the electron, once noise and pileup are taken into account.
To ensure isolation, the scalar sum of the \pt of all tracker and
calorimeter objects in a cone of radius
$\Delta R=\sqrt{\smash[b]{(\Delta \phi)^2 + (\Delta \eta)^2}}<0.3$
where $\phi$ is azimuthal angle in radians, must be less than 3\% of the electron \pt.
Only well-measured
tracks that are consistent with originating from the same vertex as the electron are included
in the isolation cone.

The muon system covers the pseudorapidity region $\abs{\eta}<2.4$.
The reconstruction of muons is optimized for high-\pt values~\cite{ZprimeFull2016}.
Measurements from the inner tracker and the outer muon system are combined.
The tracker track must have a transverse impact parameter $d_{xy} < 2\unit{mm}$
with respect to the primary vertex.
The longitudinal distance of the tracker track from the primary vertex must
be less than 5\unit{mm}.
Each muon track is required to have at least one hit in the pixel
detector, at least six tracker layer hits, and segments with hits in two or more muon detector stations. Since segments are
typically
in consecutive layers separated by thick layers of steel,
the latter requirement significantly reduces the amount of hadronic punch-through~\cite{Chatrchyan:2012xi}.
In order to suppress muons with mismeasured \pt, an additional requirement $\sigma_{\pt}/\pt<0.3$ is applied, where
$\sigma_{\pt}$ is the \pt uncertainty from the muon track reconstruction.
Muon isolation requires that the scalar \pt sum of all tracks originating from the interaction
vertex within a $\Delta R<0.3$ cone around its direction, excluding the muon itself, is less than 10\% of the muon's \pt.

This analysis uses $\eta$- and \pt-dependent scale factors to correct the simulated event
yields for the measured differences in the trigger, reconstruction, and identification
efficiencies as compared with data.

For each event, hadronic jets are clustered from the reconstructed particle-flow objects with the infrared
and collinear safe anti-\kt jet
clustering algorithm~\cite{Cacciari:2008gp, Cacciari:2011ma} using a distance parameter of $R = 0.4$.
Jet momentum is determined as the vectorial sum of all particle momenta in the jet, and is found from simulation to be, on average, within
5 to 10\% of the
true momentum over the whole \pt spectrum and detector acceptance.
Additional proton-proton interactions within the same or nearby bunch crossings
can contribute
additional tracks and calorimetric energy depositions, increasing the apparent jet
momentum.
To mitigate this effect, tracks identified as originating from
pileup vertices
(contributions from additional pp interactions in the same or nearby bunch crossings)
are discarded.
Jet energy corrections are derived from simulation
studies so that the average measured response of jets becomes identical to that of particle level jets. In situ
measurements of the momentum balance in dijet,
photon+jet, \PZ{}+jet, and multijet events are used to determine any residual differences between the jet energy scale in
data and in simulation, and appropriate
corrections are made~\cite{cms_collaboration_jet_2016}. Additional selection criteria are applied to each jet to remove
jets potentially dominated by instrumental
effects or reconstruction failures.

\section{Backgrounds}

Because the final state contains two opposite-sign leptons, the main background
originates from
Drell--Yan (DY) production.
These events are simulated using the NLO generator \MGvATNLO v2.2.2~\cite{Alwall:2014hca,Frederix:2012ps}
with up to two final-state partons,
and are hadronized with \PYTHIA 8.212.
Because large mass lepton pairs associated with large hadronic activity are a small fraction
of all DY events, additional high mass simulations are used to ensure sufficient statistics
to evaluate this background.
The cross section is normalized to next-to-next-to-leading-order (NNLO) QCD and
next-to-leading-order (NLO) EWK using a di-lepton mass-dependent
K factor~\cite{ZprimeFull2016,GAVIN20112388}.

Another important background is the \ttbar process, which yields two prompt leptons
if both top quarks decay semi-leptonically.
This becomes the dominant background in the medium-mass validation region (VR), as will
be described in Section~\ref{sec:selection}.
It is simulated with \POWHEG 2.0~\cite{powheg_1, powheg_2, powheg_3, Re:2010bp, Frixione:2007nw}
in combination with \PYTHIA~8.212 for hadronization.
Dedicated high-mass \ttbar samples are used to provide sufficient number of events to estimate
the background contribution to the high-mass signal region, and are simulated with the
same method as used for the main \ttbar sample.
The cross section is normalized to NNLO~\cite{PhysRevLett.110.252004}.
Production of single top quarks in association with a \PW-boson
or in the $t$-channel production are simulated
with \POWHEG. The single top $s$-channel is simulated with \MGvATNLO.

Several multi-boson processes can also yield at least two leptons.
However, because they have
small cross sections,
they are sub-dominant backgrounds in this search.
The $\PW\PW$, $\PW\PZ$, and $\PZ\PZ$ processes with at least two leptons are simulated
with \POWHEG and \MGvATNLO.
Triple-boson processes are neglected because of their small cross sections.

The $\PW$+jets production can contribute to the signal region when jets are
misidentified as electrons.
In contrast, the misidentification of muons is negligible.
The $\PW$+jets production is simulated to leading order with
\MGvATNLO v2.2.2~\cite{Alwall:2014hca,Frederix:2012ps}.
In addition,  high-\HT \PW samples are also used,
where \HT is defined as the sum of jet transverse momenta.
The $\PW$+jets
cross section is normalized to NNLO by application of a K-factor of 1.21~\cite{GAVIN20112388}.

The background samples are simulated
using the NNPDF2.3 or NNPDF3.0~\cite{ball_parton_2015} PDF sets with tune CUETP8M1
(in 2016) and the NNPDF3.1 (in 2017) PDF set with CP5 (2017) for the underlying event,
respectively~\cite{GEN-17-001}.
All generated events are processed through a full simulation of the CMS detector
based on
\GEANTfour~\cite{Agostinelli:2002hh}, a trigger emulation, and the event reconstruction chain.
All simulated event samples are normalized to the integrated luminosity of the recorded data.
The simulation
of pileup
is included in all events by superimposing simulated minimum bias interactions.
For the data set used, the average number of interactions per bunch crossing after selection
is about 23 in 2016 and about 30 in 2017, with maximum values of 55 and 70,
respectively.

\section{Event selection and validation regions}
\label{sec:selection}

Events with electrons in the final state were selected using a trigger that requires an
electromagnetic cluster (electron or photon) with transverse momenta larger than 175\GeV (for 2016)
and 200\GeV (for 2017). 
Events with muons in the final state were triggered with
the single-muon trigger with a minimum \pt of 50\GeV.
The electron (muon) efficiency plateaus at 98.5\% (98\%)~\cite{wprime}
as a function of \pt
above a turn-on region and is independent of
$\eta, \phi$, and pileup.

\begin{figure}[htb]
  \centering
  \includegraphics[width=0.5\textwidth]{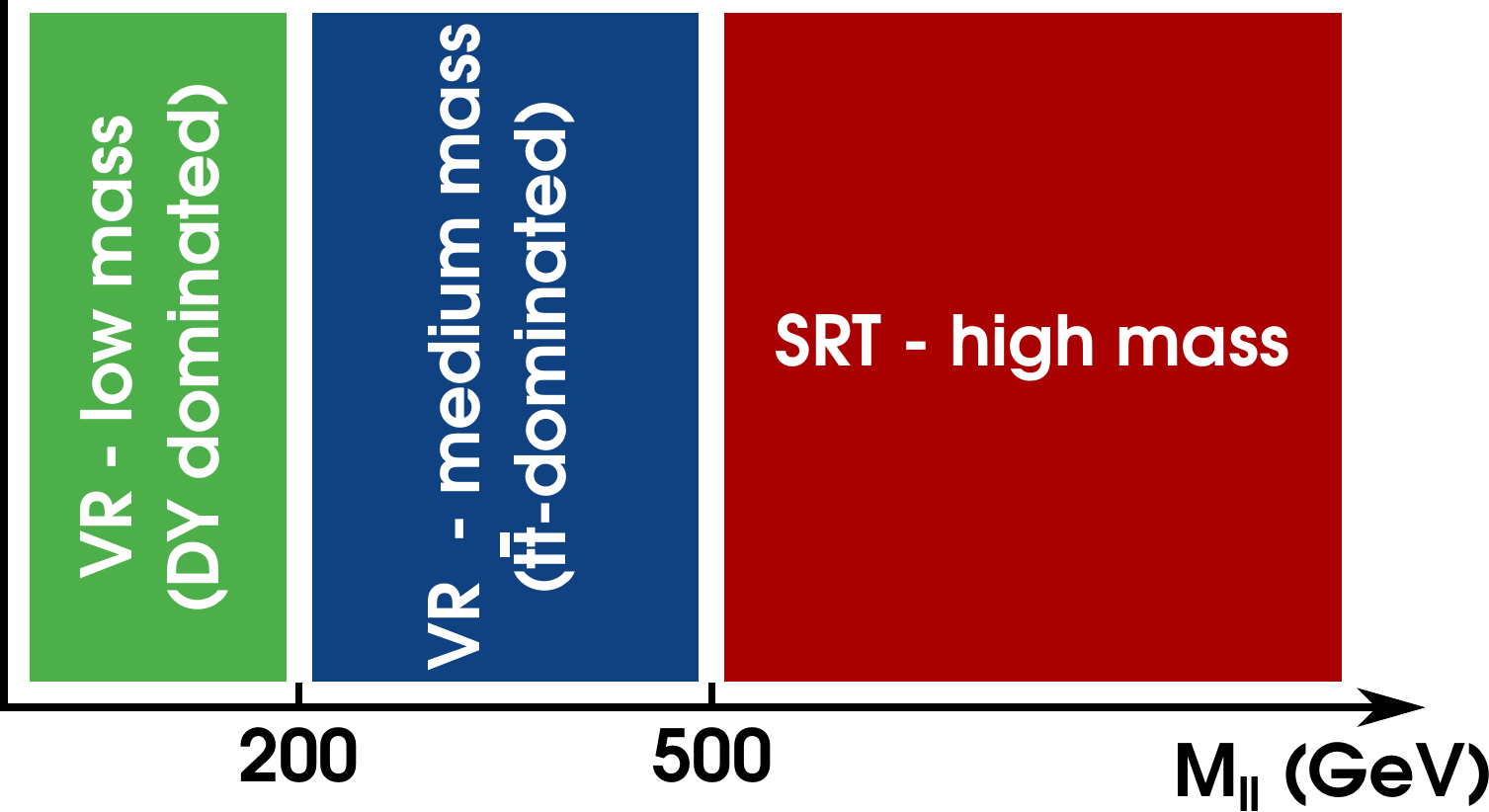}
  \caption{Definition of the two validation regions (VR) and the high-mass signal region (SRT).}
  \label{fig:regions}
\end{figure}

Final state particles (electrons, muons and jets) are reconstructed as described in
Section~\ref{sec:objects}.
The two highest \pt (leading) leptons are selected along with the two leading jets.
The \pt thresholds for the leading lepton
are 230\GeV (electron) and 53\GeV (muon), chosen to be above the respective trigger turn-on regions.
The subleading electrons or muons
are required to have an offline \pt of at least 35 and 25\GeV, respectively.
Jets are required to have a minimum \pt of 50\GeV.
Events with more than two same-flavor leptons are rejected, to reduce background from diboson production.

The majority of the DY background is suppressed by requiring $M_{\ell\ell} > 500\GeV$,
constituting the \textit{high-mass signal region (SRT)}.
Two VRs are then defined, illustrated in Fig.~\ref{fig:regions},
to verify agreement between data and simulation in different mass regions.
The \textit{low-mass VR} with $M_{\ell\ell} < 200\GeV$ serves as the VR for DY, which
is the dominant background for this final state.
The subleading \ttbar background tends to populate the
\textit{medium-mass VR}, which is defined by $200\GeV < M_{\ell\ell} < 500\GeV$.
While the low-mass VR is nearly signal free, the medium-mass VR
potentially contains a very small fraction
of signal events.
The signal contamination in the medium VR is far below 1\% for the interesting mass range of \mlstar above
3.8\TeV.

\begin{figure}[hbt]
\centering
      \includegraphics[width=0.49\textwidth]{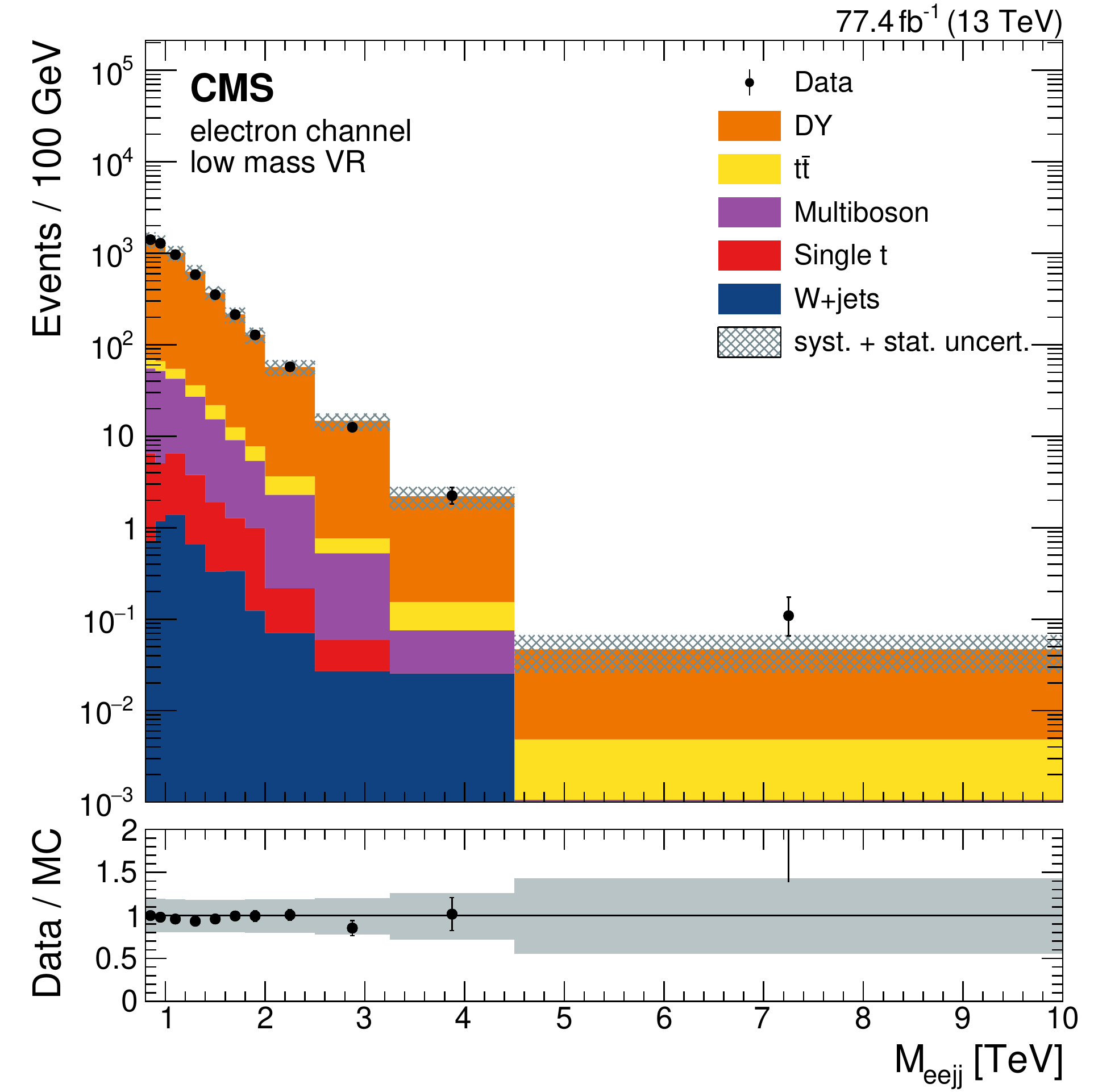}
      \includegraphics[width=0.49\textwidth]{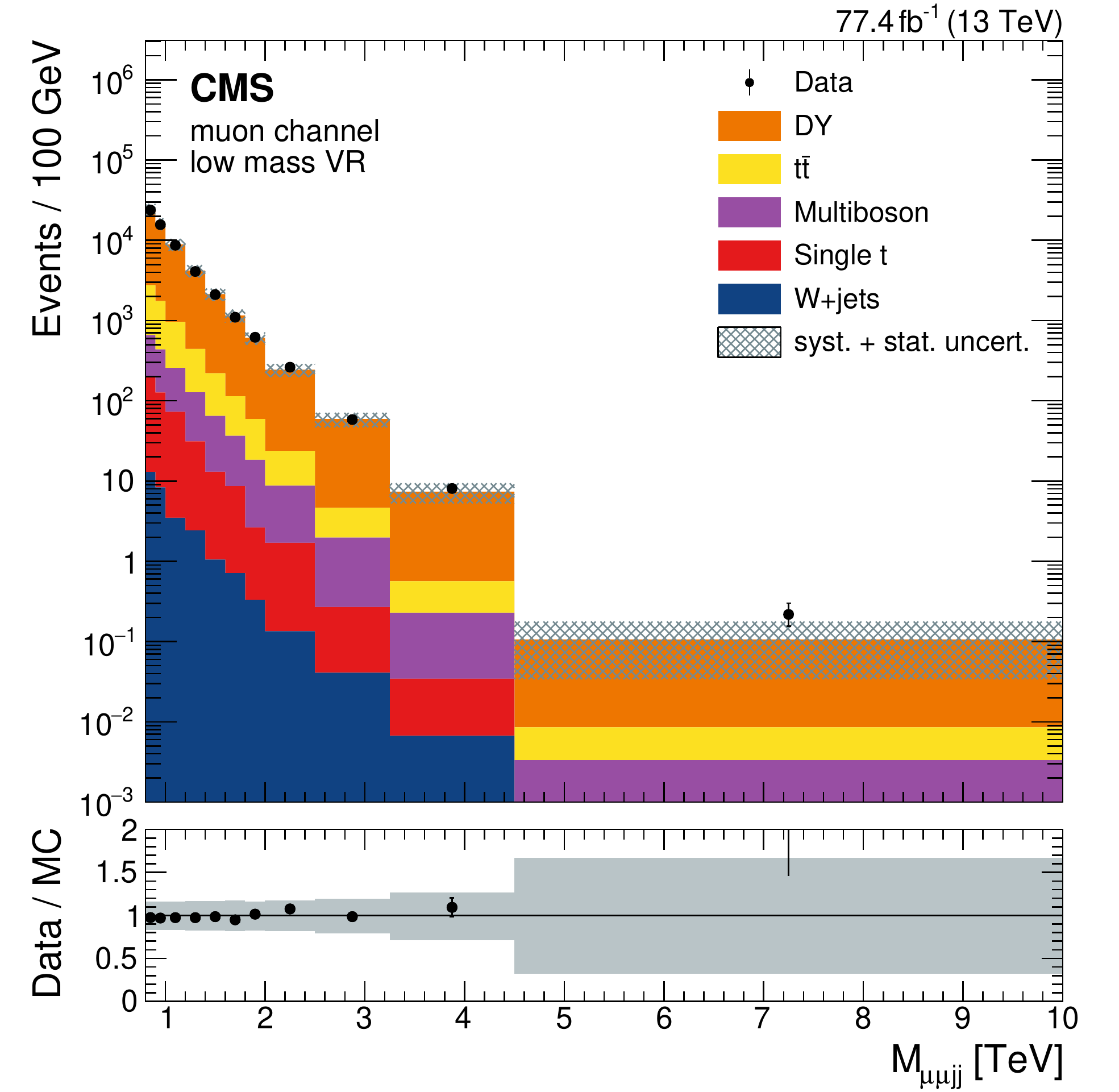}
   \caption{
Event distributions as a function of the four-body invariant $M_{\lljj}$ mass for the electron (left)
and muon (right) channels, for the low-mass validation
region defined by $M_{\ell\ell}<200\GeV$.
The lower panels show the ratio of data to the simulated SM background, with the shaded band representing
the uncertainty.
}
\label{fig:low-massCR}
\end{figure}

\begin{figure}[hbt]
\centering
      \includegraphics[width=0.49\textwidth]{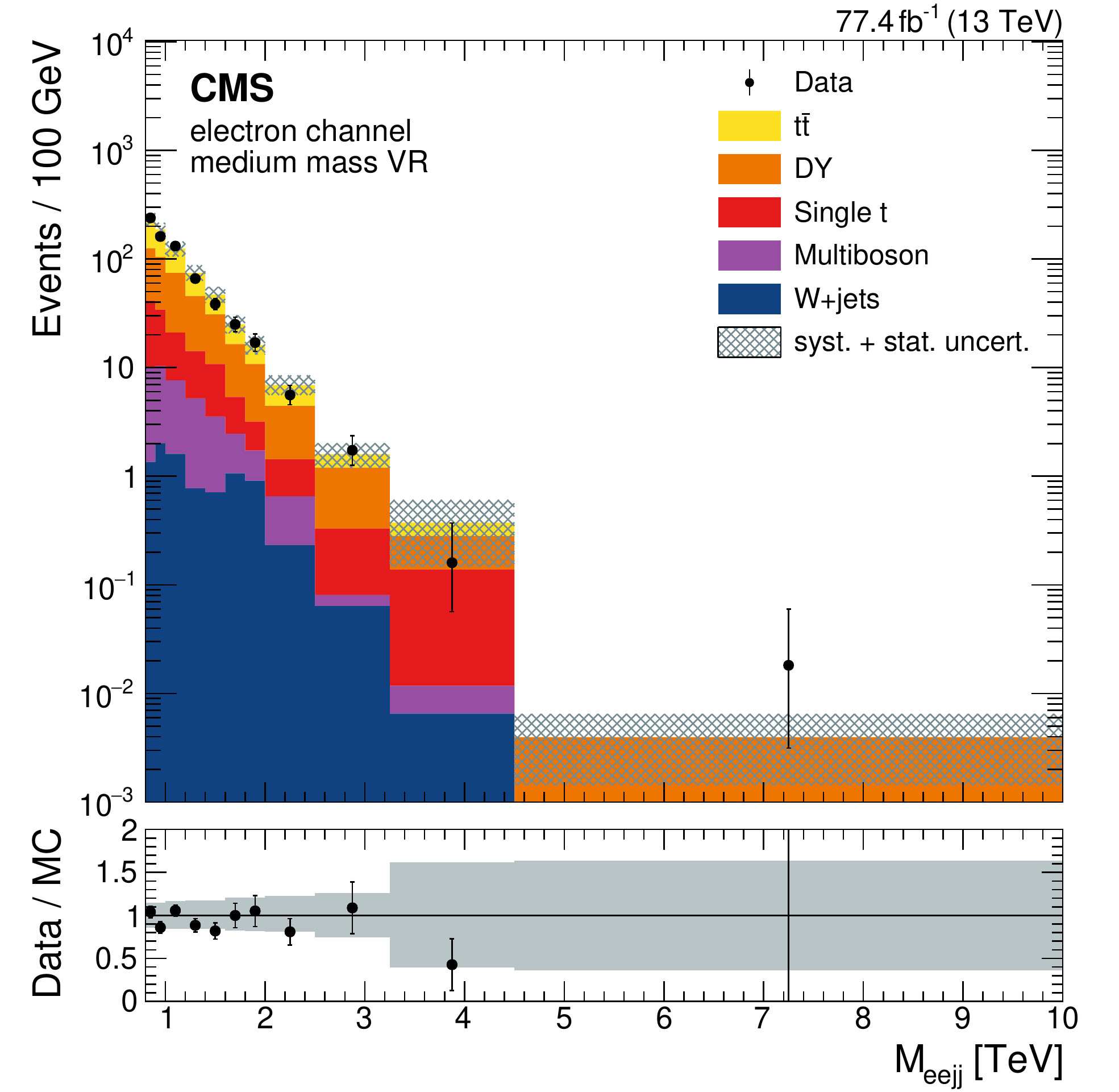}
      \includegraphics[width=0.49\textwidth]{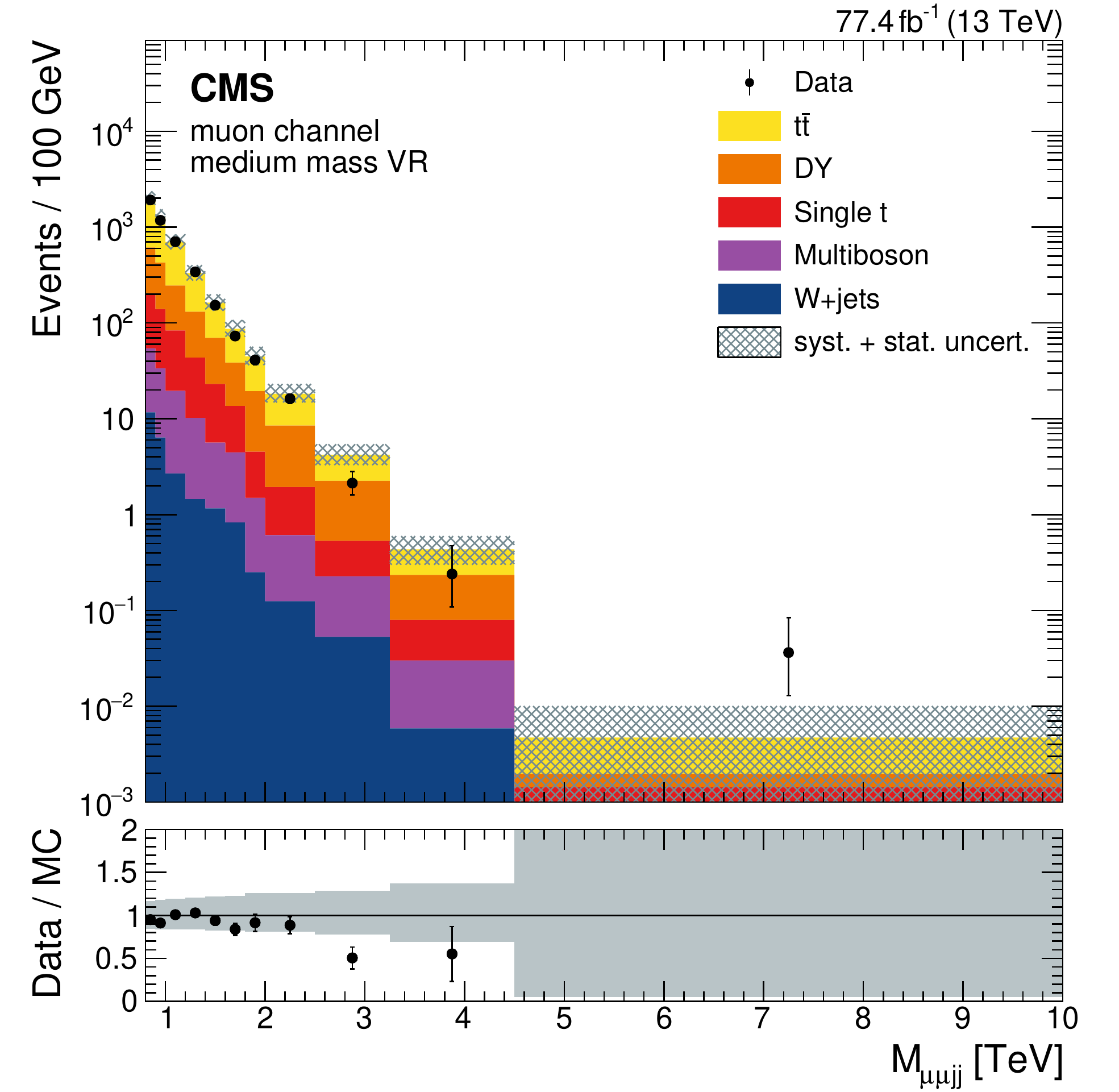}
   \caption{
Event distribution as a function of the four-body invariant mass $M_{\lljj}$
for the electron (left) and muon (right) channels, for the medium-mass
validation region defined by $200<M_{\ell\ell}<500\GeV$.
The lower panel shows the ratio of data to the simulated SM background, with the shaded band representing
the uncertainty.
}
   \label{fig:medium-massCR}
\end{figure}

The invariant mass of the combination of both selected leptons and the two leading jets,
$M_{\lljj}$,
is used as the discriminating variable as it provides the best separation power between signal and background.
The distributions of other possible variables were
studied, but their use yielded
lower signal significance.
The event distributions as functions of the four-body mass
for the low-mass and medium-mass validation regions
are shown in Figs.~\ref{fig:low-massCR} and \ref{fig:medium-massCR}, respectively.
The good agreement between experimental and simulated data justifies estimating
the background expectation from simulation.


\begin{figure}[hbt]
 \centering
  \includegraphics[width=0.49\textwidth]{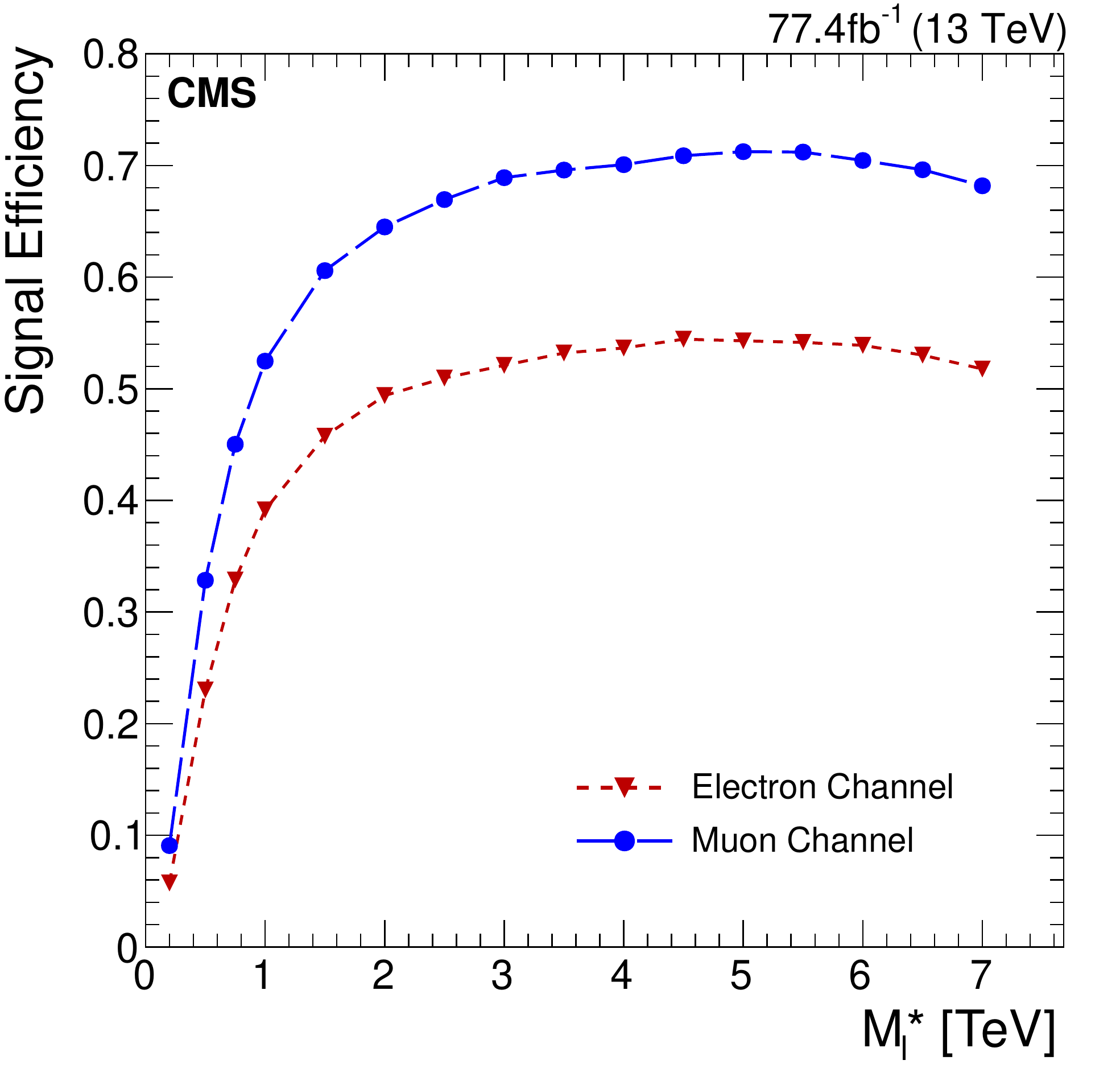}
  \caption{Signal efficiency after all selections are applied, as
a function of the excited lepton mass \mlstar, based on simulated events.
 }
 \label{fig:efficiency}
\end{figure}

The signal efficiency, defined as the product of acceptance and detection
efficiency of the fraction of reconstructed 2$\ell$2j
events,
resulting from these event selections is shown as a function
of the simulated excited lepton mass in Fig.~\ref{fig:efficiency}.
The uncertainties in the figure are only statistical uncertainties.
The systematic uncertainties on the product of
acceptance and efficiency are between 6 and 8\%.
The efficiency
reaches a plateau of about 55\% for the electron channel and 73\% for the muon channel.
For lower masses, in particular for $\mlstar< 1\TeV$, the acceptance is reduced by the offline
\pt thresholds
on the objects
but this mass region is not of primary interest given previous exclusion limits.
For masses above 6\TeV, the efficiency starts to drop slightly because of a
growing fraction of events produced
off-shell. At the given center-of-mass energy of the LHC, the available energy for generating such heavy
particles is limited.

\section{Systematic uncertainties}
\label{sec:systematics}

Systematic uncertainties in this analysis can affect the overall normalization and also
the shape of the distributions.
Uncertainties in the energy scale of different physics objects
are taken into account by analyzing the shape of the four-body mass distributions, with
the energy scale shifted up and down by $1\,\sigma$.
The electron energy scale uncertainty is estimated to be 0.2\% in the barrel and 0.3\% in the
endcap~\cite{electron_reco_2015}.
For muons,
the momentum scale is determined using the generalized endpoint
method~\cite{MuonPerformance13TeV} and applied as a function of the
$\eta$ and $\phi$ of the muon. The studies indicate the absence of a significant curvature bias,
within an uncertainty of 0.02/\TeVns in the central region
and up to 0.1/TeV in the more forward regions.
For jets, the uncertainties associated with the jet energy correction are
used~\cite{cms_collaboration_jet_2016}.

The simulated energy resolution is better than that measured in data.
To account for this, jet energy resolution corrections are applied to jets
and their uncertainties are considered as a systematic effect~\cite{cms_collaboration_jet_2016}.
For electrons, the momentum is smeared by 1.2 and 2.4\% for barrel and endcaps, respectively~\cite{electron_reco_2015}.
For muons, the momentum is smeared by 5\%~\cite{MuonPerformance13TeV}.
These three uncertainties change the shape of the background.

There are systematic uncertainties on the scale factors applied to correct for the differences
in the triggering, reconstruction and identification efficiencies between simulation and data.
These uncertainties are typically a few percent
and are taken into account as
normalization uncertainty~\cite{electron_reco_2015,MuonPerformance13TeV}.
The integrated luminosity of the data collected has an uncertainty of
2.5~\cite{CMS-PAS-LUM-17-001} and 2.3\%~\cite{CMS-PAS-LUM-17-004}
for the 2016 and 2017 data sets, respectively.
The uncertainty due to the modeling of pileup is found to be less than 10\%
and is by
derived by 
calculating the pileup for the cases where the total inelastic cross section 
is increased and decreased by 5\% from the nominal value~\cite{Sirunyan:2018nqx}.
It is treated as an uncertainty in the background shape.

The uncertainty associated with the choice of PDFs
affects the cross section of the simulated samples and is taken into
account by
following an approach outlined by the PDF4LHC recipe~\cite{butterworth_pdf4lhc_2016}.
The PDF set used to calculate the variations is NNPDF3.0, for background and signal
samples in 2016, and NNPDF3.1, in 2017.
Overall, the variation is found to range from 5\% to 60\% for an excited lepton mass 
from 1\TeV to 6\TeV, respectively,
as taken from
LHAPDF6.2~\cite{buckley_lhapdf6:_2015}.

The renormalization and factorization scales provide a handle
to estimate the uncertainty due to missing higher orders.
Both scales are varied simultaneously by a factor 2 up and down, resulting in different
event-by-event weights and an
uncertainty of 10 to 30\% between 1 and 6\TeV.

In the signal region, the limited number of simulated background events results in a
statistical uncertainty of up to 40\% in
the background prediction.

Production of $\PW$+jets final states can contribute to the background when a jet is
misidentified as a lepton,
albeit at a very small level.
A conservative systematic uncertainty of 100\% is applied to the $\PW$+jets background
contribution
and reflects the uncertainty in the lepton misidentification rate in the simulation as compared
with data.

For the SM background, the dominant sources of uncertainties are the Monte--Carlo (MC)
sample sizes and the PDF
uncertainty.

\section{Results}
\label{sec:results}

\begin{figure}[hbtp]
 \includegraphics[width=0.49\textwidth]{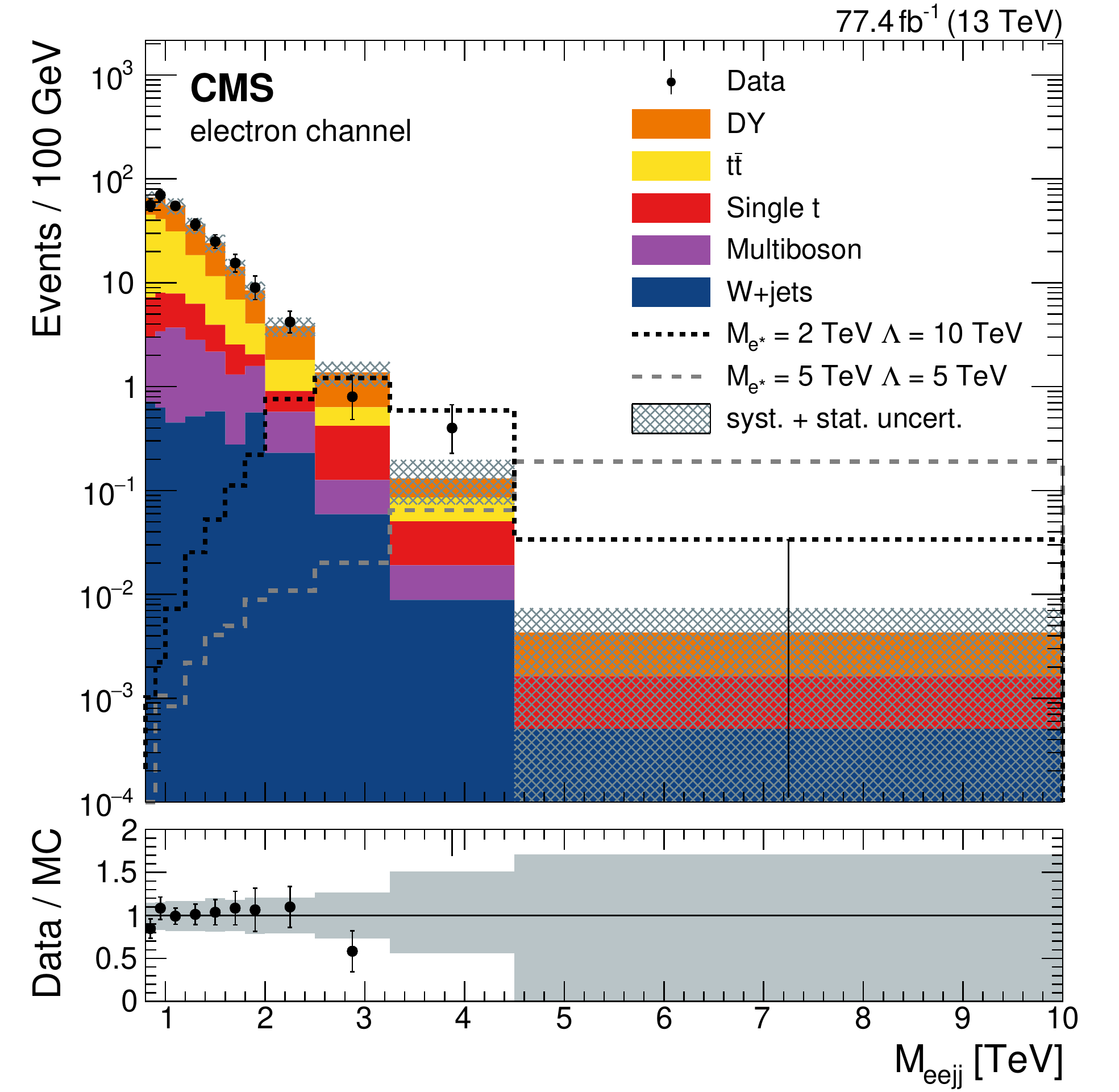}
 \includegraphics[width=0.49\textwidth]{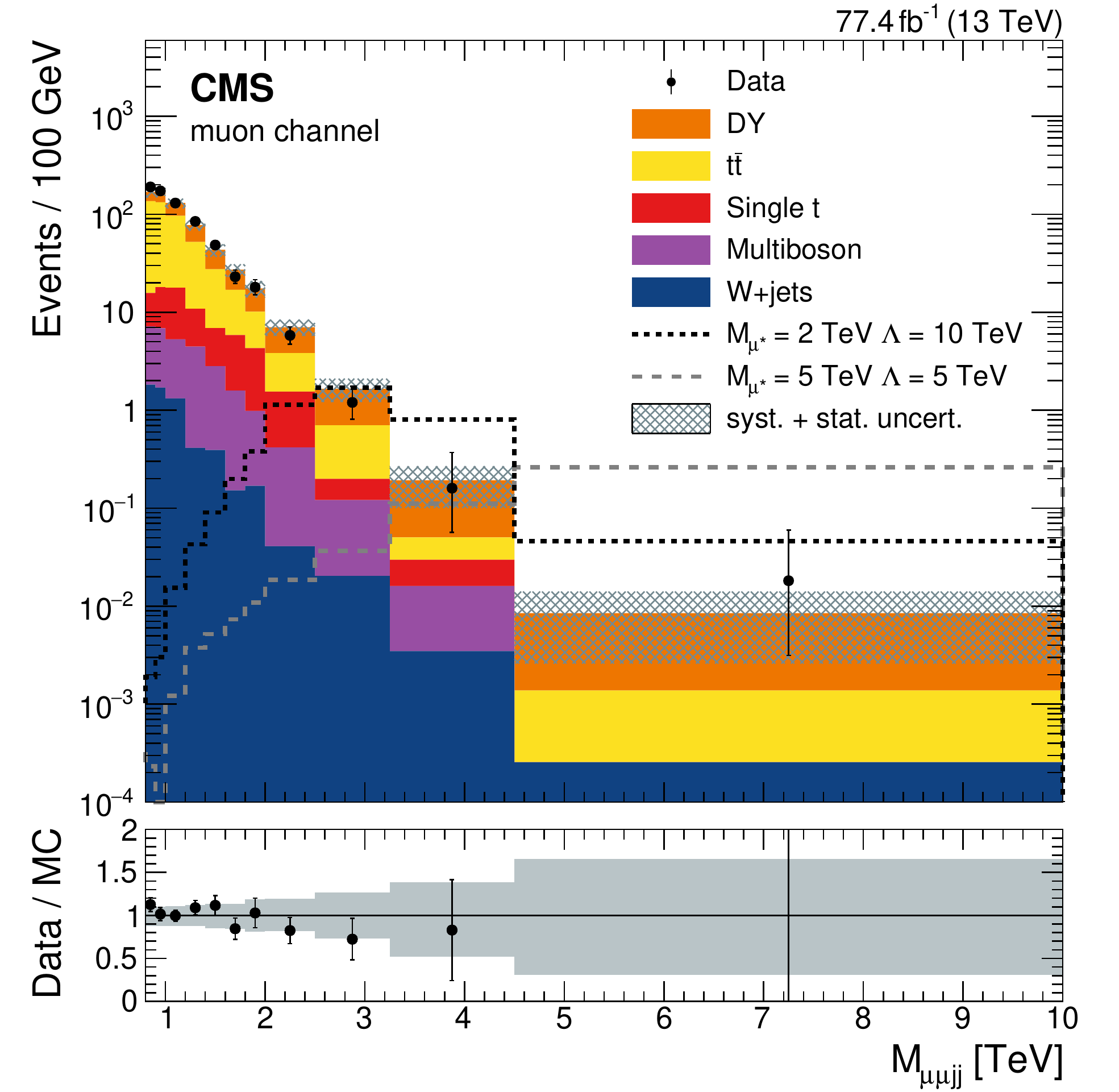}
 \caption{Distribution of the two-lepton two-jet invariant mass in the signal region
($M_{\ell\ell} > 500\GeV$) for the electron (left) and muon (right) channels.
The example signal shape for two excited lepton masses is indicated as a gray line with the
parameters given in the legend and for the benchmark case where the couplings $f$ and $f'$
are set to unity.
The lower panels show the ratio of data to simulation with the total uncertainty in gray.
}
\label{fig:final_distribution}
\end{figure}

Event distributions for the signal regions as a function of the four-body invariant mass
are shown for the electron and muon channels in
Fig.~\ref{fig:final_distribution}.
The final distribution is dominated by \ttbar background in the low-mass region and DY
background in
the high
mass region despite being strongly suppressed by the dilepton invariant mass cut.
The expected background in the highest mass bin, of more than 5\TeV width, is far less than one
event.
The analysis would not benefit from reducing the background further at the cost of
signal efficiency.
The event yields in data compared to the total expected SM backgrounds are given
in Table~\ref{tab:event_yields}
for a number of bins of the discriminating four-body $2\ell2j$ mass distribution.
Also shown are the expectations for potential \lstar signals with two different sets
of model parameter values: a mass \mlstar of 2\TeV
and compositeness scale $\Lambda$ of 10\TeV; and $\mlstar = \Lambda = 5\TeV$,
which roughly corresponds to the maximum sensitivity of this analysis.

\begin{table}[htbp]
   \topcaption{
Observed event yields in bins of four-body mass
compared to the expected SM background, for the $2e2j$ and $2\mu2j$ final
states.
Also shown are the expected event yields for two simulated signal samples with the given masses and couplings.
All yields are given in bins of the discriminating four-body mass ($2\ell2j$) distribution, with lower and
upper value for each
bin given in units of GeV.
Systematic uncertainties, as described in
the text, are shown.
}
\label{tab:event_yields}
\cmsTable{
\begin{tabular}{lccccc}
\hline
   & \multicolumn{5}{c}{Bins in $2\ell2j$ mass ({\GeVns})} \\
   & 800--1600  & 1600--2500  & 2500--3250   & 3250--4500 & 4500--10000 \\ [\cmsTabSkip]
\hline
Data in $2e2j$        & 359                & 70           & 6             & 5           & 0 \\
SM prediction         & $362^{+58}_{-61}$    & $65 \pm 12$    & $10.3^{+2.7}_{-2.8}$  & $1.6^{+0.8}_{-0.7}$ & $0.24^{+0.17}_{-0.24}$ \\
                      &                      &                &                 &               &                    \\
$f = f' = 1$  &                      &                &                 &               &                    \\
$M_{\estar} = 2\TeV, \Lambda = 10\TeV$
                      & $0.17^{+0.01}_{-0.02}$ & $4.5 \pm 0.3$ & $9.1 \pm 0.6$ & $7.4 \pm 0.5$ & $1.9 \pm 0.2$ \\
$M_{\estar} = 5\TeV, \Lambda = 5\TeV$
                      & $0.015^{+0.002}_{-0.003}$ & $0.08 \pm 0.01$ & $0.15 \pm 0.01$ & $0.81^{+0.07}_{-0.06}$ & $10.4 \pm 0.7$ \\
$f = f' = 0.1$            &                      &                &                 &               &                    \\
$M_{\estar} = 2\TeV, \Lambda = 10\TeV$ & $0.53 \pm 0.05$ & $13.6^{+1.0}_{-0.9}$ & $27.8^{+1.7}_{-1.8}$ & $22.6 \pm 1.6$ & $5.7 \pm 0.5$ \\
$M_{\estar} = 5\TeV, \Lambda = 5\TeV$  & $0.016^{+0.002}_{-0.003}$ & $0.09 \pm 0.01$ & $0.16 \pm 0.01$ & $0.85^{+0.08}_{-0.07}$ & $10.9 \pm 0.8$ \\
[\cmsTabSkip]
\hline
Data in $2\mu2j$      & 889                & 111              & 9                 & 2           & 1 \\
SM prediction   & $842^{+91}_{-102}$  & $125^{+18}_{-19}$  & $12.4\pm3.3$         & $2.4^{+0.9}_{-1.2}$ & $0.47^{+0.31}_{-0.32}$ \\
                      &                      &                &                 &               &                    \\
$f = f' = 1$  &                      &                &                 &               &                    \\
$M_{\mustar} = 2\TeV, \Lambda = 10\TeV$ & $0.30\pm0.02$ & $6.9^{+0.3}_{-0.2}$ & $12.7^{+0.5}_{-0.4}$ & $10.1^{+0.6}_{-0.4}$ & $2.5 \pm 0.2$ \\
$M_{\mustar} = 5\TeV, \Lambda = 5\TeV$ & $0.021\pm0.003$ & $0.13 \pm 0.01$ & $0.27^{+0.02}_{-0.01}$ & $1.4 \pm 0.1$ & $14.4^{+1.2}_{-0.4}$ \\
$ f = f' = 0.1$  &                      &                &                 &               &                    \\
$M_{\mustar} = 2\TeV, \Lambda = 10\TeV$ & $0.92^{+0.06}_{-0.05}$ & $20.9^{+0.9}_{-0.7}$ & $38.8^{+1.6}_{-1.1}$ & $30.8^{+1.8}_{-1.1}$ & $7.7^{+0.7}_{-0.5}$ \\
$M_{\mustar} = 5\TeV, \Lambda = 5\TeV$  & $0.021 \pm 0.003$ & $0.14 \pm 0.01$ & $0.29^{+0.02}_{-0.01}$ & $1.5 \pm 0.1$ & $15.1^{+1.2}_{-0.4}$ \\
\hline
\end{tabular}
}
\end{table}

No indication of a signal is observed.
The limits on the excited lepton mass, \mlstar, are calculated using the Bayesian
method~\cite{PDG2018}
with
a uniform positive prior probability distribution for the signal cross section. Systematic
uncertainties in the expected signal and background yields are included either via nuisance
parameters with log-normal prior distributions or with the shape of the distribution
included through the use of a binned likelihood.

\begin{figure}[hbt]
\centering
  \includegraphics[width=0.49\textwidth]{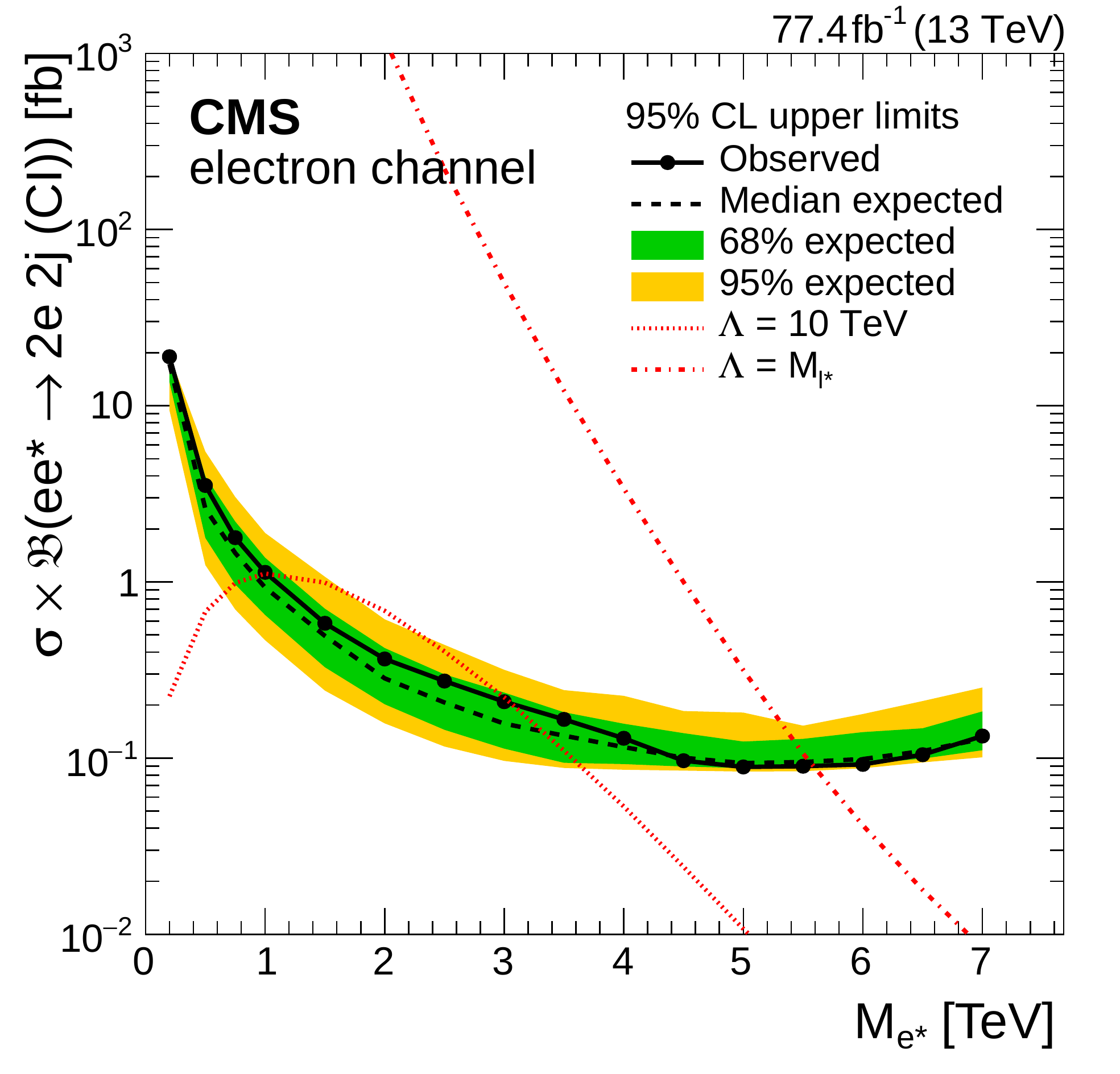}
  \includegraphics[width=0.49\textwidth]{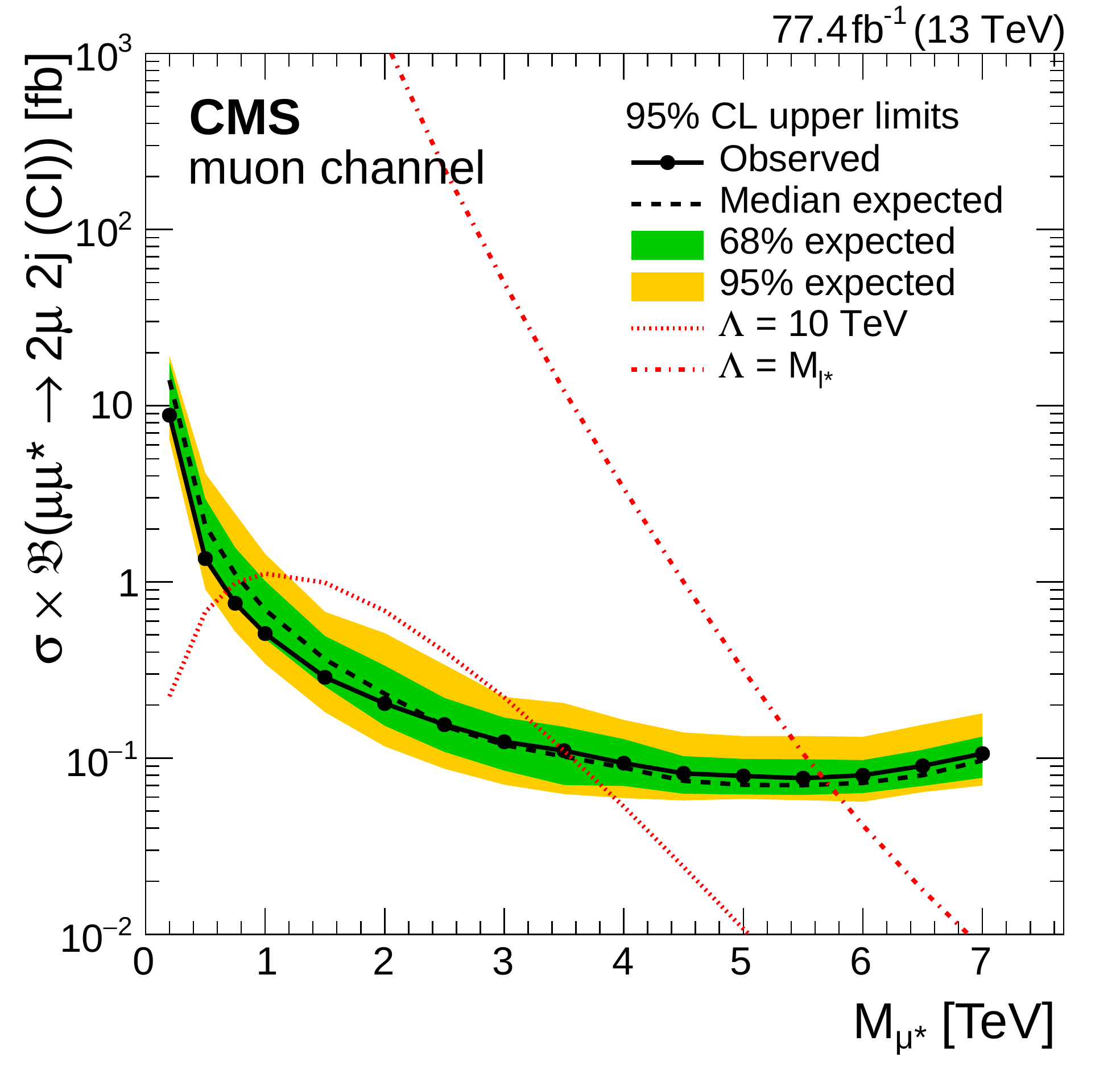}
  \caption{Limits at 95\% \CL on the
product of the production cross section and branching fraction for
$\ell\lstar\to\lljj$,
as a function of the excited lepton mass $\mlstar$,
for the electron (left)
and muon (right) channels. The expectation from the model is represented
for $\abs{f} = \abs{f'} = 1$
by two cases, $\Lambda = 10\TeV$, and
$\Lambda = \mlstar$.
}
\label{fig:limit_xsec}
\end{figure}

Limits at 95\% confidence level (\CL)
on the product of cross section and branching fraction assuming
the proper branching fraction for \mlstar=$\Lambda$ and $f=f'=1$
are shown in Fig.~\ref{fig:limit_xsec},
and range from 10 to 0.1\unit{fb}, depending on the mass of the excited lepton.

This analysis excludes
excited electrons and excited muons at 95\% \CL for
values of \mlstar below 5.6 and 5.7\TeV, respectively, assuming
that $\Lambda$ is equal to the mass of the excited leptons.
These are the best limits to date on excited electrons and muons.
At low masses, the sensitivity is determined by the acceptance and analysis selection. At very high mass,
the sensitivity becomes limited by the cross section.
The sensitivity to the maximum \lstar mass is not affected by the coupling strength.

\begin{figure}[hbt]
\centering
  \includegraphics[width=0.49\textwidth]{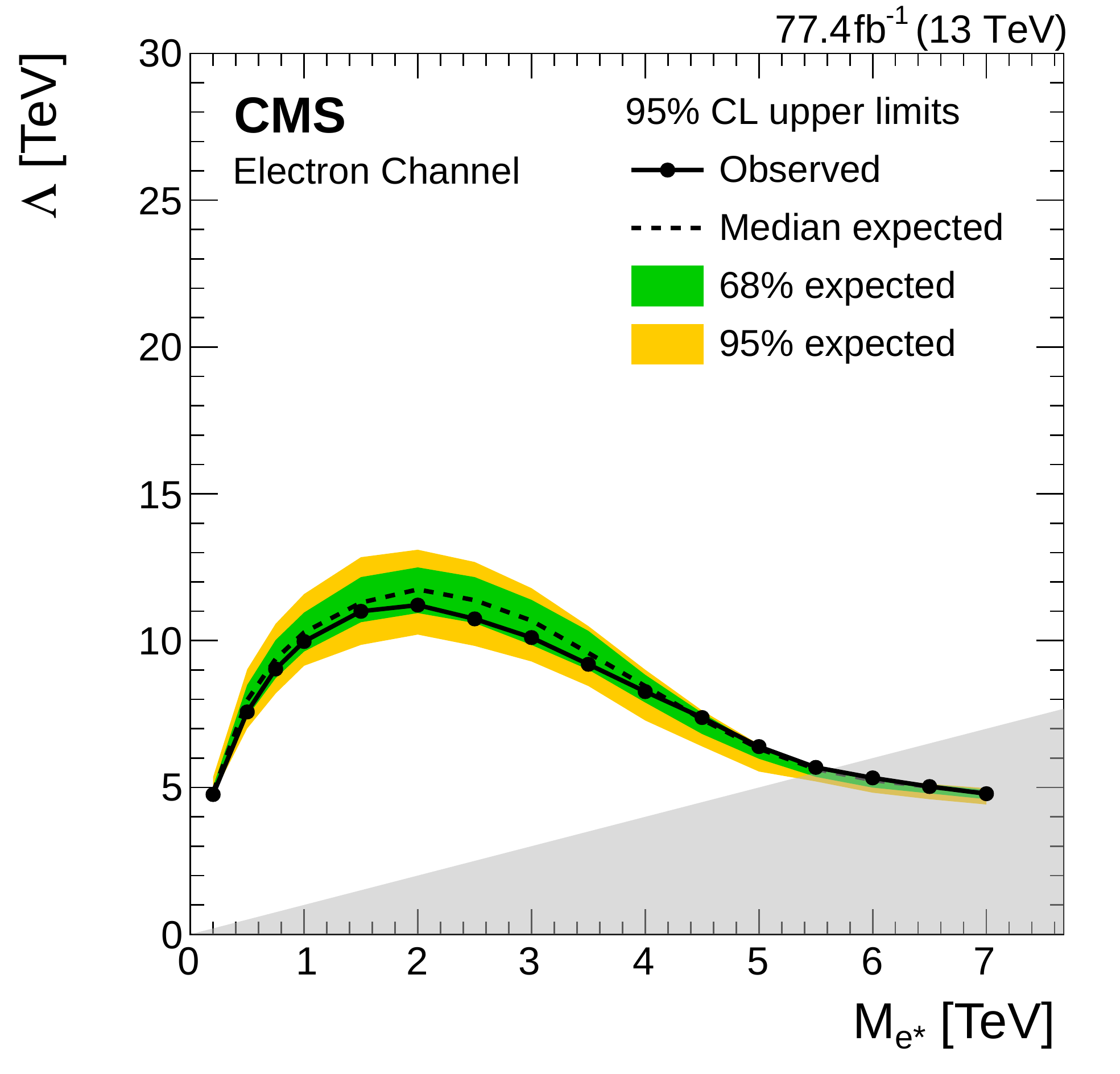}
  \includegraphics[width=0.49\textwidth]{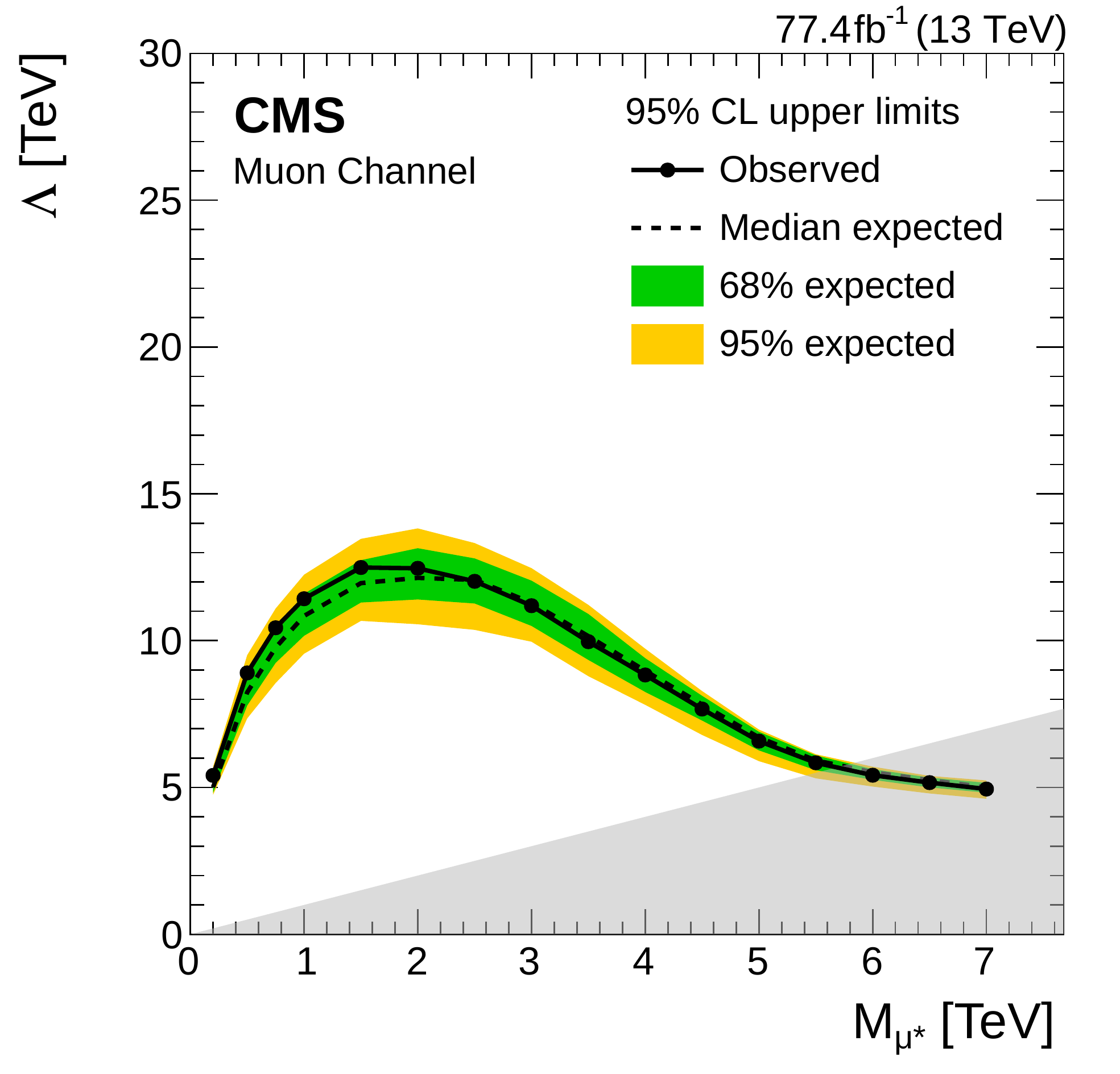}
  \caption{Limits on the compositeness scale $\Lambda$ for the electron (left) and muon (right) channels,
as a function of the mass of the excited lepton,
 for the benchmark case where the GI couplings $\abs{f}$ and $\abs{f'}$ are set to one.
 The model is not valid in the hatched area.
 }
 \label{fig:limit_lambda_f1}
\end{figure}

\begin{figure}[hbt]
\centering
  \includegraphics[width=0.49\textwidth]{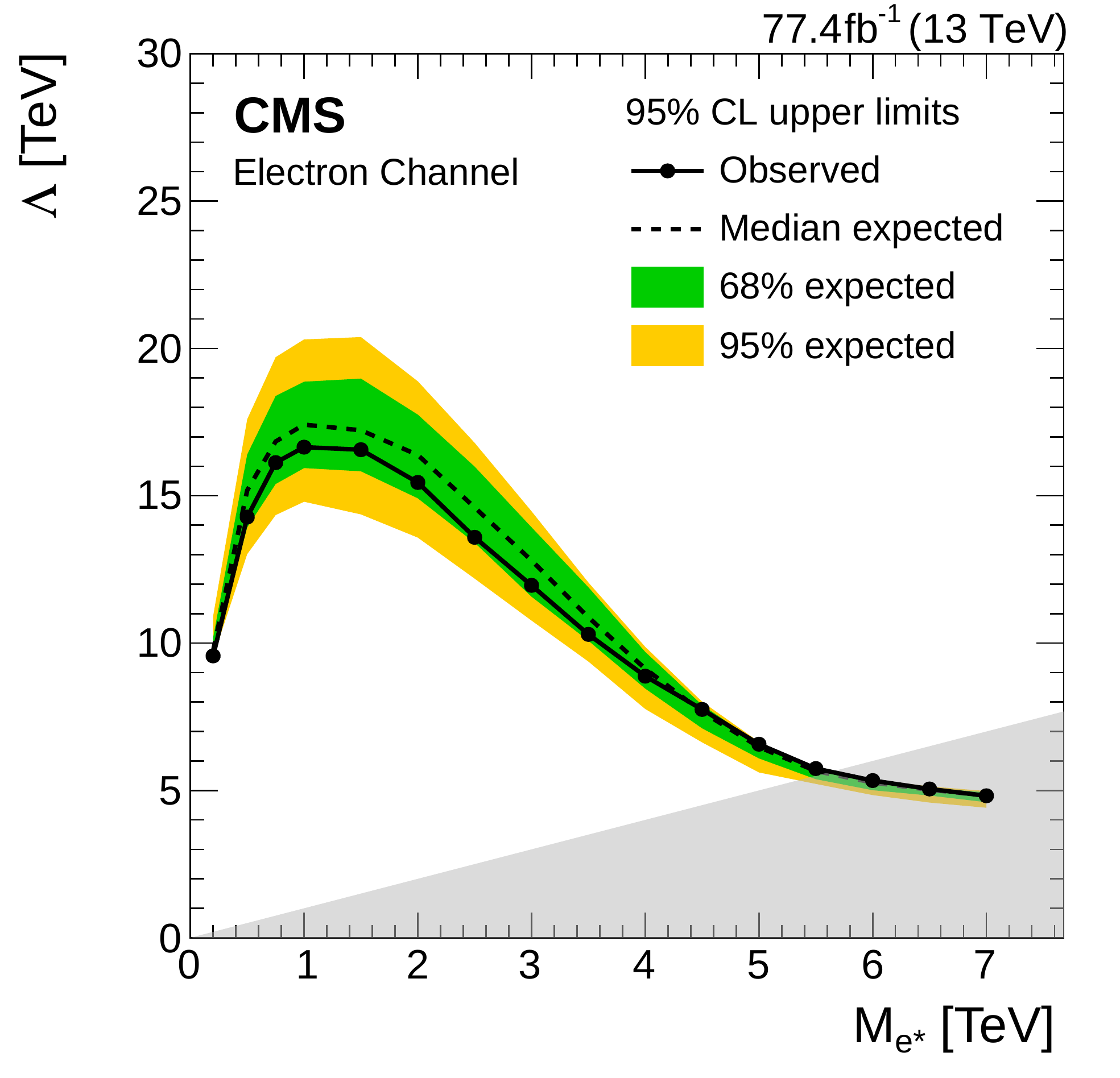}
  \includegraphics[width=0.49\textwidth]{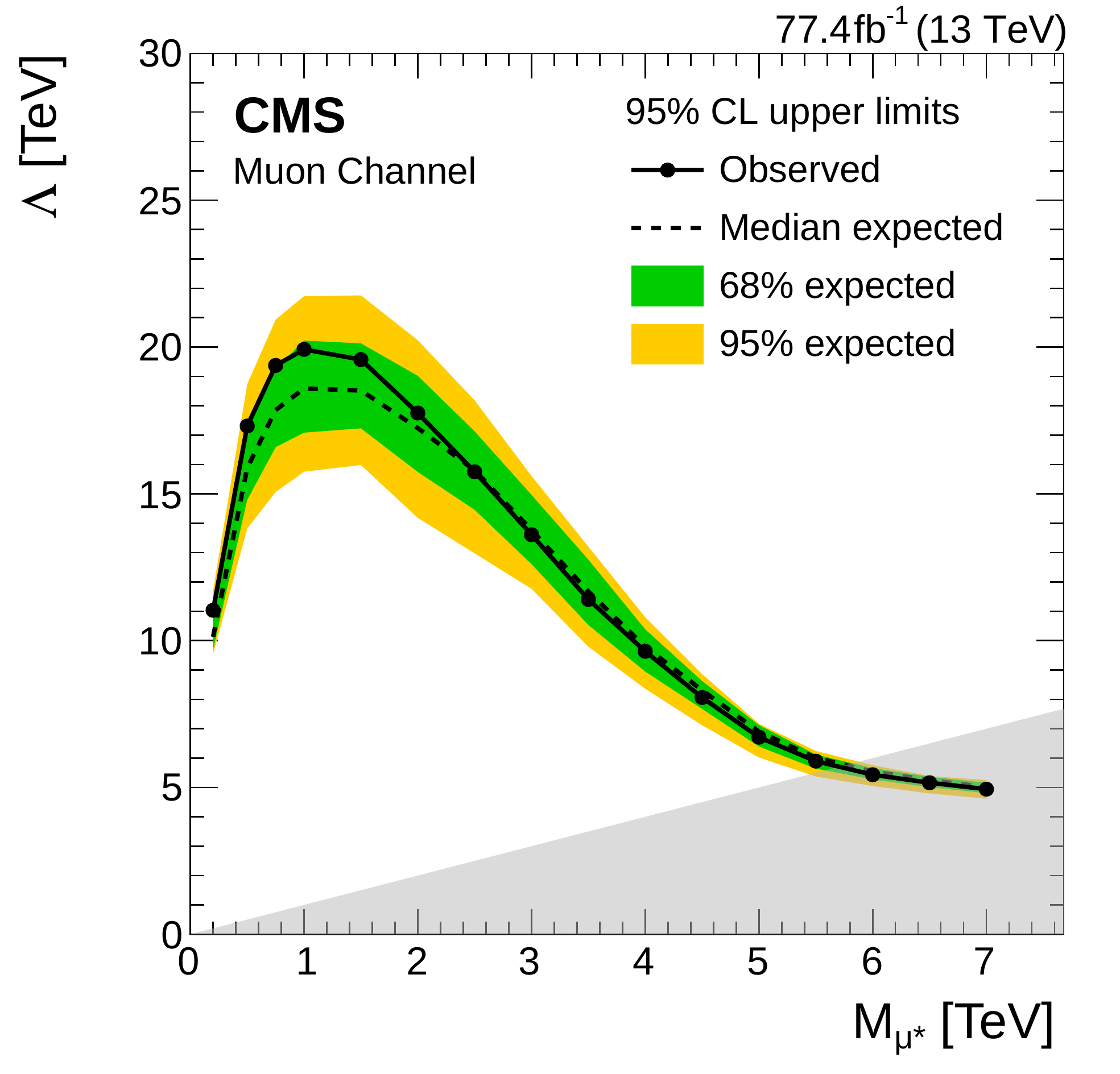}
  \caption{Limits on the compositeness scale $\Lambda$ for the electron (left) and muon (right) channels,
as a function of the mass of the excited lepton,
 for the case where the GI couplings $\abs{f}$ and $\abs{f'}$ are set to 0.1.
 The model is not valid in the hatched area.
 }
 \label{fig:limit_lambda_f01}
  \includegraphics[width=0.49\textwidth]{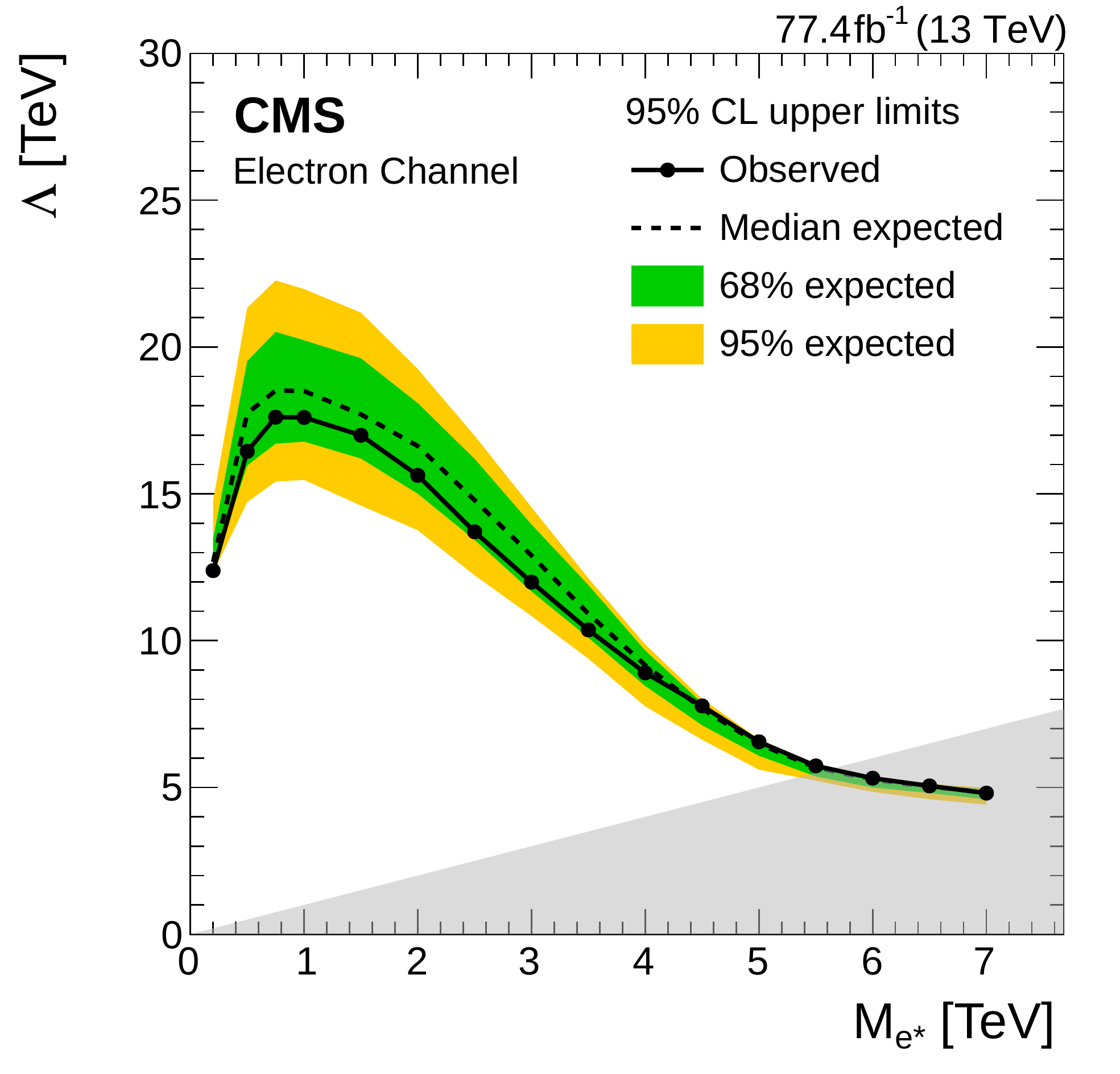}
  \includegraphics[width=0.49\textwidth]{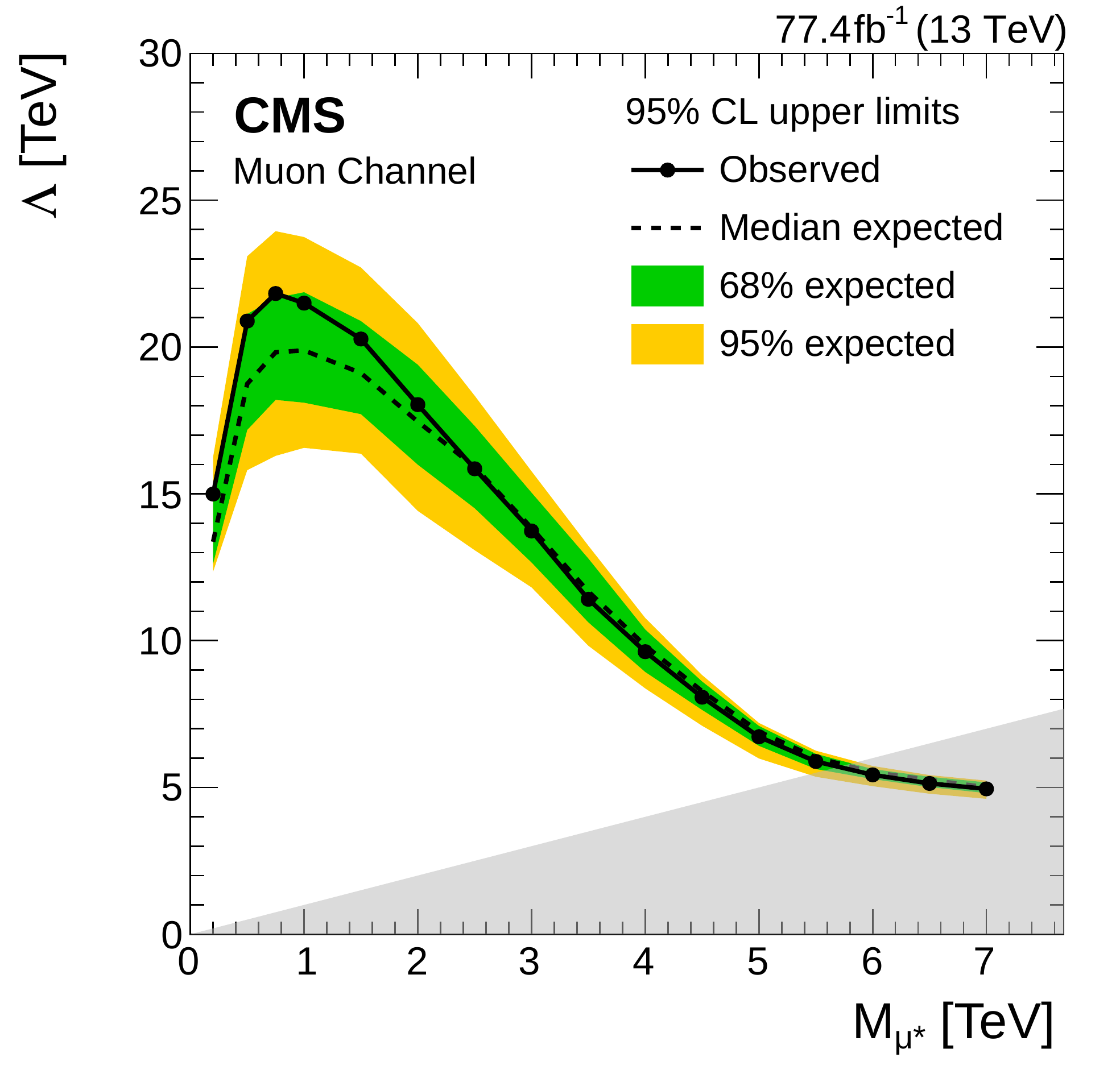}
  \caption{Limits on the compositeness scale $\Lambda$ for the electron (left) and muon (right) channels,
as a function of the mass of the excited lepton,
 for the benchmark case where the GI couplings $f$ and $f'$ vanish.
 The model is not valid in the hatched area.
 }
 \label{fig:limit_lambda_f0}
\end{figure}

The cross section limit can be re-evaluated in terms of the compositeness scale $\Lambda$ as a function
of
the excited lepton mass.
This sensitivity does depend on the coupling strength.
Figures~\ref{fig:limit_lambda_f1}--\ref{fig:limit_lambda_f0}
show the variations in the compositeness
scale $\Lambda$ for the gauge couplings $\abs{f}$ and $\abs{f'}$ equal to unity,
for $\abs{f}$ and
$\abs{f'}$ equal to 0.1, and for $f$ and $f'$ equal to zero, respectively.
Compositeness scales below 5\TeV are excluded for all
investigated \lstar masses.
Maximum sensitivity to $\Lambda$ is reached at low masses, as is
typical
for excited lepton searches. The exact mass for this peak sensitivity
depends on the coupling scenario,
decreasing with weaker GI couplings, and can easily be understood from Fig.~\ref{fig:br}.
Detailed numbers are given in
Table~\ref{tab:SummaryLimits}.
In the case of $f = f' = 1$, compositeness scales up to
11\TeV (\estar) and 12\TeV (\mustar)
are excluded, with the maximum sensitivity for \lstar masses around 2\TeV.
With values for $f = f' = 0.1$, the decay width to a CI increases, yielding a higher
$\Lambda$ sensitivity of 17\TeV (\estar) and 19\TeV (\mustar), respectively,
for \lstar masses around 1.5\TeV.
Assuming zero GI couplings, the observed
$\Lambda$ sensitivity increases to 18\TeV (\estar) and 22\TeV (\mustar), respectively,
for \lstar masses around 1\TeV.
While the expected sensitivities in the electron and muon channels are comparable,
the observed muon channel sensitivity is up to one standard deviation
higher due to fluctuations in the data.
For \mlstar above 2\TeV these are the best limits to date.
In the scenario $ f = -f'$, where \llg has no sensitivity,
this CI channel provides
the best test of high compositeness scales.

\clearpage
\section{Summary}

A search for excited leptons decaying via a contact interaction to final states of
two electrons or two muons and
two resolved jets has been performed.
This channel
complements other searches for excited leptons.
It has greatest sensitivity at
large values of the excited lepton mass \mlstar.
The data for this analysis were recorded with the CMS detector in the years 2016
and 2017, corresponding to a total integrated luminosity of 77.4\fbinv
of proton-proton collisions at a center-of-mass energy of 13\TeV.

No significant deviations from SM expectations are observed in the signal region and 95\% exclusion
limits have been set.
Excited electrons (muons)  up to masses of
$M_{\estar} = 5.6\TeV$ ($M_{\mustar} = 5.7\TeV$) are excluded with the usual assumption of
$\mlstar = \Lambda$.
These are the best limits to date.
The limit was also re-evaluated in terms of the substructure scale $\Lambda$,
leading to limits
of $\Lambda = 11$ and 12\TeV for excited electrons and muons, respectively,
for mass values around 2\TeV
and couplings of unity.
When studying weaker gauge couplings, the limit on the maximum \mlstar does not change,
but the larger cross section increases
the $\Lambda$ sensitivity at lower masses. For couplings around zero,
where the \llg decay has no sensitivity,
limits around 20\TeV for the compositeness scale $\Lambda$
are achieved.

\begin{table}[hbtp]
\centering
\topcaption{Summary of the observed (expected) limits on \lstar mass, assuming \mlstar = $\Lambda$,
for the cases $f = f'$ and $f = -f'$.
The limits evaluated in terms of the compositeness scale $\Lambda$ are shown in the right column.
}
\label{tab:SummaryLimits}
\begin{tabular}{l l c c l}
\hline
\multirow{2}{*}{Search channel} & Coupling & \multicolumn{2}{c}{ \mlstar $= \Lambda$, values in \TeV} & Limit on $\Lambda$, in \TeV \\
                                & strength & $f = f' $                 & $f  = -f' $                                 &
\\
\hline
$\Pe\Pe^* \, \, \to  2\Pe 2j$
   & $f = 1$   & 5.6 (5.6)    & 5.6 (5.6)  & 11 (11) for $\mlstar \approx 2\TeV$ \\
   & $f = 0.1$ & 5.6 (5.6)    & 5.6 (5.6)  & 17 (18) for $\mlstar \approx 1.5\TeV$ \\
   & $f = 0$   & 5.6 (5.6)    & 5.6 (5.6)  & 18 (19) for $\mlstar \approx 1\TeV$  \\
$\Pgm\Pgm^*\to 2\Pgm 2j$
   & $f = 1$   & 5.7 (5.7)    & 5.7 (5.7)  & 12 (12) for $\mlstar \approx 2\TeV$ \\
   & $f = 0.1$ & 5.7 (5.7)    & 5.7 (5.7)  & 19 (19) for $\mlstar \approx 1.5\TeV$ \\
   & $f = 0$   & 5.7 (5.7)    & 5.7 (5.7)  & 22 (20) for $\mlstar \approx 1\TeV$   \\
\hline
\end{tabular}
\end{table}

\begin{acknowledgments}
We congratulate our colleagues in the CERN accelerator departments for the excellent performance of the LHC and thank the technical and administrative staffs at CERN and at other CMS institutes for their contributions to the success of the CMS effort. In addition, we gratefully acknowledge the computing centers and personnel of the Worldwide LHC Computing Grid for delivering so effectively the computing infrastructure essential to our analyses. Finally, we acknowledge the enduring support for the construction and operation of the LHC and the CMS detector provided by the following funding agencies: BMBWF and FWF (Austria); FNRS and FWO (Belgium); CNPq, CAPES, FAPERJ, FAPERGS, and FAPESP (Brazil); MES (Bulgaria); CERN; CAS, MoST, and NSFC (China); COLCIENCIAS (Colombia); MSES and CSF (Croatia); RPF (Cyprus); SENESCYT (Ecuador); MoER, ERC IUT, PUT and ERDF (Estonia); Academy of Finland, MEC, and HIP (Finland); CEA and CNRS/IN2P3 (France); BMBF, DFG, and HGF (Germany); GSRT (Greece); NKFIA (Hungary); DAE and DST (India); IPM (Iran); SFI (Ireland); INFN (Italy); MSIP and NRF (Republic of Korea); MES (Latvia); LAS (Lithuania); MOE and UM (Malaysia); BUAP, CINVESTAV, CONACYT, LNS, SEP, and UASLP-FAI (Mexico); MOS (Montenegro); MBIE (New Zealand); PAEC (Pakistan); MSHE and NSC (Poland); FCT (Portugal); JINR (Dubna); MON, RosAtom, RAS, RFBR, and NRC KI (Russia); MESTD (Serbia); SEIDI, CPAN, PCTI, and FEDER (Spain); MOSTR (Sri Lanka); Swiss Funding Agencies (Switzerland); MST (Taipei); ThEPCenter, IPST, STAR, and NSTDA (Thailand); TUBITAK and TAEK (Turkey); NASU (Ukraine); STFC (United Kingdom); DOE and NSF (USA).

\hyphenation{Rachada-pisek} Individuals have received support from the Marie-Curie program and the European Research Council and Horizon 2020 Grant, contract Nos.\ 675440, 752730, and 765710 (European Union); the Leventis Foundation; the A.P.\ Sloan Foundation; the Alexander von Humboldt Foundation; the Belgian Federal Science Policy Office; the Fonds pour la Formation \`a la Recherche dans l'Industrie et dans l'Agriculture (FRIA-Belgium); the Agentschap voor Innovatie door Wetenschap en Technologie (IWT-Belgium); the F.R.S.-FNRS and FWO (Belgium) under the ``Excellence of Science -- EOS" -- be.h project n.\ 30820817; the Beijing Municipal Science \& Technology Commission, No. Z191100007219010; the Ministry of Education, Youth and Sports (MEYS) of the Czech Republic; the Deutsche Forschungsgemeinschaft (DFG) under Germany’s Excellence Strategy -- EXC 2121 ``Quantum Universe" -- 390833306; the Lend\"ulet (``Momentum") Program and the J\'anos Bolyai Research Scholarship of the Hungarian Academy of Sciences, the New National Excellence Program \'UNKP, the NKFIA research grants 123842, 123959, 124845, 124850, 125105, 128713, 128786, and 129058 (Hungary); the Council of Science and Industrial Research, India; the HOMING PLUS program of the Foundation for Polish Science, cofinanced from European Union, Regional Development Fund, the Mobility Plus program of the Ministry of Science and Higher Education, the National Science Center (Poland), contracts Harmonia 2014/14/M/ST2/00428, Opus 2014/13/B/ST2/02543, 2014/15/B/ST2/03998, and 2015/19/B/ST2/02861, Sonata-bis 2012/07/E/ST2/01406; the National Priorities Research Program by Qatar National Research Fund; the Ministry of Science and Education, grant no. 14.W03.31.0026 (Russia); the Programa Estatal de Fomento de la Investigaci{\'o}n Cient{\'i}fica y T{\'e}cnica de Excelencia Mar\'{\i}a de Maeztu, grant MDM-2015-0509 and the Programa Severo Ochoa del Principado de Asturias; the Thalis and Aristeia programs cofinanced by EU-ESF and the Greek NSRF; the Rachadapisek Sompot Fund for Postdoctoral Fellowship, Chulalongkorn University and the Chulalongkorn Academic into Its 2nd Century Project Advancement Project (Thailand); the Kavli Foundation; the Nvidia Corporation; the SuperMicro Corporation; the Welch Foundation, contract C-1845; and the Weston Havens Foundation (USA).
\end{acknowledgments}

\bibliography{auto_generated}

\cleardoublepage \appendix\section{The CMS Collaboration \label{app:collab}}\begin{sloppypar}\hyphenpenalty=5000\widowpenalty=500\clubpenalty=5000\vskip\cmsinstskip
\textbf{Yerevan Physics Institute, Yerevan, Armenia}\\*[0pt]
A.M.~Sirunyan$^{\textrm{\dag}}$, A.~Tumasyan
\vskip\cmsinstskip
\textbf{Institut f\"{u}r Hochenergiephysik, Wien, Austria}\\*[0pt]
W.~Adam, F.~Ambrogi, T.~Bergauer, J.~Brandstetter, M.~Dragicevic, J.~Er\"{o}, A.~Escalante~Del~Valle, M.~Flechl, R.~Fr\"{u}hwirth\cmsAuthorMark{1}, M.~Jeitler\cmsAuthorMark{1}, N.~Krammer, I.~Kr\"{a}tschmer, D.~Liko, T.~Madlener, I.~Mikulec, N.~Rad, J.~Schieck\cmsAuthorMark{1}, R.~Sch\"{o}fbeck, M.~Spanring, D.~Spitzbart, W.~Waltenberger, C.-E.~Wulz\cmsAuthorMark{1}, M.~Zarucki
% ...
\vskip\cmsinstskip
\dag: Deceased\\
1:  Also at Vienna University of Technology, Vienna, Austria\\
% ...
\end{sloppypar}
%%% END EDITABLE REGION %%%
\end{document}